\documentclass[a4paper,10pt]{article}

\usepackage[latin1]{inputenc}

\usepackage[fleqn]{amsmath} 

\usepackage{amsfonts}
\usepackage{amssymb}
\usepackage[english]{babel}
\usepackage{comment}
\usepackage{graphicx}
\usepackage{rotating}
\usepackage{booktabs}

{\left\lbrace\begin{array}{@{}l@{}}}%
{\end{array}\right.}

\DeclareMathOperator{\DFT}{DFT}

\DeclareMathOperator{\RDFT}{RDFT}
\DeclareMathOperator{\CDFT}{CDFT}
\DeclareMathOperator{\DCT}{DCT}
\DeclareMathOperator{\DST}{DST}

\begin{document}

\title{a class of AM-QFT algorithms for power-of-two FFT}

%\title{8 variants of QFT algorithm for power-of-two FFT}

\author{Lorenzo Pasquini\footnote{\em Falconara Marittima (AN), Italy, pasquini.paper@gmail.com}}

%\authorrunning{Pasquini} % if too long for running head

%\date{December 2012}

\maketitle

\begin{abstract} 

This paper proposes a class of power-of-two FFT (Fast Fourier Transform) algorithms, called AM-QFT algorithms, that contains the improved QFT  (Quick Fourier Transform), an algorithm recently published, as a special case.
The main idea is to apply the Amplitude Modulation Double Sideband - Suppressed Carrier (AM DSB-SC) to convert odd-indices signals into even-indices signals, and to insert this elaboration into the improved QFT algorithm, substituting the multiplication by secant function. 
The 8 variants of this class are obtained by re-elaboration of the AM DSB-SC idea, and by means of duality.
As a result the 8 variants have both the same computational cost and the same memory requirements than improved QFT.
Differently, comparing this class of 8 variants of AM-QFT algorithm with the split-radix 3add/3mul (one of the most performing FFT approach appeared in the literature), we obtain the same number of additions and  multiplications, but employing half of the trigonometric constants.
This makes the proposed FFT algorithms interesting and useful for fixed-point implementations.
Some of these variants show advantages versus the improved QFT.
In fact one of this variant slightly enhances the numerical accuracy of improved QFT, while other four variants use trigonometric constants that are faster to compute in `on the fly' implementations.

\end{abstract} 

%\keywords{Discrete Fourier Transform \and DFT \and Quick Fourier Transform \and DCT \and DST \and QFT \and FFT \and Split-Radix \and computational complexity}

%%%%%%%%%%%%%%%%%%%

\section{Introduction} \label{sec:introduction}

In many engineering and theoretical applications we need to compute $\DFT$ (Discrete Fourier Transform).
Direct calculation of $\DFT$ is computationally demanding  ($cost(N) \sim O(N^2)$), where $N$ is the signal length. 
Many FFT (Fast Fourier Transform) algorithms exist \cite{Duhamel_Vetterli_1990} to reduce such a cost to $cost(N) \sim N \cdot \log (N)$. 
In power-of-two FFT context, the radix-2 is the simplest and most famous of these.
The split-radix \cite{Duhamel_Hollmann_1984}, \cite{Martens_1984}, \cite{Vetterli_Nussbaumer_1984} (of whom many variants exists:  \cite{Bouguezel_2007}, \cite{Gopinath_1989}, \cite{Kamar_Elcherif_1989}, \cite{Stasinski_1991}, \cite{Yavne_1968},) perhaps is the best compromise between computational cost, simplicity, and memory requirements.
A class of scaled algorithms \cite{Bernstein_2007}, \cite{Johnson_Frigo_2007}, \cite{Lundy_Van_Buskirk_2007} reaches the minimum computational cost, in a computational model that evaluates efficiency with required flops (multiplications plus additions on floating-point values).
The improved QFT algorithm \cite{Pasquini_2013} is a recently appeared algorithm, that has a computational cost identical to split-radix 3add/3mul, but using half trigonometric constants.
In this paper we propose a class (8 variants) of algorithms for power-of-two FFT that we can obtain re-elaborating the approach leading to the improved QFT. 
The new idea consists in using the AM DSB-SC modulation (instead of multiplication by secant function) in the improved QFT context, to convert odd-indices signals into even-indices signals.
As a second step we can also re-elaborate the AM DSB-SC idea in different ways (as using duality), to transform odd-indices signals into even-indices ones, maintaining the advantages of improved QFT, that is itself a variant (the 4th) of this class of AM-QFT algorithms.
We describe this class of algorithms using the new `language' (definitions of new concepts, and a new notation) already used in \cite{Pasquini_2013}.

Here is the outline of the paper. First, in sect.\ref{sec:language}, we briefly describe (and further develop) the approach used in \cite{Pasquini_2013} to delineate algorithms.
Then, in sect.\ref{sec:transformations} we analyze the main employed elaborations (transformations and decompositions) shared by the algorithms proposed in this paper. 
Sect.\ref{sec:improved_QFT} shows a brief resume of the improved QFT.
Sect.\ref{sec:le_varianti_della_QFT},\ref{sec:6_algorithms} describe respectively the ideas behind the innovative developed algorithms, and their structure.
Then, in sect.\ref{sec:costo_computazionale}, we discuss the memory requirement, the computational cost and the accuracy of QFT variants, also highlighting advantages, disadvantages and possible applications. 
Finally, sect.\ref{sec:finale} summarizes the results of this paper.

%%%%%%%%%%%%%%%%%%%%%%%%%%%%%%%%%%%%%%%%%%%%%%%

\section{The new approach used to describe FFT algorithms} \label{sec:language}

In our opinion the use of language introduced in \cite{Pasquini_2013} represents significant advantages in order to approach many kinds of FFT algorithms.
Some of them (such as the compactness of description of algorithms) become even more relevant in this paper than in \cite{Pasquini_2013}.
In fact, by virtue of the new language, the 129 signals required to describe the 24 distinct functions used by the 8 AM-QFT algorithms, can be classified in only 18 different signal types (that have to be handled in different ways).
Moreover many of these signal types have already been created in \cite{Pasquini_2013}, and the others are dual to the ones used in \cite{Pasquini_2013}.
Thus we use the same approach used in \cite{Pasquini_2013}, combining it with a new abstract description of algorithms tecnicque too, in order to increase the compactness of exposition.

This approach is made of many ingredients, that we briefly sum up:
\begin{itemize}

\item 
assignment of a signal type to each signal.
In order to properly characterize the signal types, we need to focus on the following signal elements: the applied transform, stored $n$ indices $sto\_n$ group, stored harmonics $sto\_k$ group, storage-size in temporal and frequency-domain $ln$ and $lk$ parameters (see sect.\ref{sec:definitions} for their definitions).
The formalization of signal types shows many advantages in development, exposition and implementations of algorithms.
Details of these advantages can be found in \cite{Pasquini_2013}.
The whole family of signal types involved in a certain algorithm, can be derived only  by a step-by-step analysis of the algorithm.

\item
use of a mnemonic notation to assign a name to each signal, and to each signal type.

\item
use of a Tab.\ref{tab:notation} to describe the characteristics of signal types, that results very helpfull in coding the algorithm in a suitable programming language. 

\item
use of Tab.\ref{tab:implementation} that describes the matching between a signal type and the array cells that store it, if the indices $n$ and $k$ are stored in growing order, and the first cell of an array has index $p=1$.

\item
splitting of signal processing applied inside each function into a sequence of basic elaborations.
Moreover, at difference with \cite{Pasquini_2013} we associate an univocal identifier to each basic elaboration in order to describe all the functions in a more compact way.
Mathematical details of used basic elaborations are listed in sect.\ref{sec:transformations} and in Tab.\ref{tab:elabora_relations_A},\ref{tab:elabora_relations_B},\ref{tab:DCT_odd_transformations},\ref{tab:DST_odd_transformations}.

\item
use of elaboration diagrams (such as Fig.\ref{fig:QFT_4_diagram},\ref{fig:QFT_2_diagram},\ref{fig:QFT_5_diagram},\ref{fig:QFT_6_diagram}) for an abstract, compact description of concatenation of basic elaborations, and involved signal types, used in each function, in a recursive version of the FFT algorithm. 
This concept is new, since it doesn't appear in \cite{Pasquini_2013}.

\item
use of decomposition tree that diagrammatically shows the global sequence of functions and signal types respectively applied to the input (root) signal and created from it (i.e. Fig.\ref{fig:QFT_2_tree}). 

\end{itemize}

%%%%%%%%%%%%%%%%%%%%%%%%%%%%%%%%%%%%

\subsection{Basic definitions} \label{sec:definitions}

Here is a brief description of used terms (for a more detailed exposition, see \cite{Pasquini_2013}).

\emph{$\DFT$, $\DCT$, $\DST$ transforms} are defined as follows:
\begin{equation} \label{equ:qft_0A}
S(k)=\DFT[s](k)= \sum_{n=0}^{N-1} s(n) \cdot e^{-i \theta \cdot n \cdot k} \quad k \in \{0,1,2,\dots,(N-1)\} 
\end{equation}
\begin{equation}  \label{equ:qft_0B}
S(k)=\DCT[s](k)= \sum_{n=0}^{\frac{N}{2}} s(n) \cdot \cos( \theta \cdot n \cdot k) \quad k \in \{0,1,2,...,(\frac{N}{2})\}
\end{equation}
\begin{equation}  \label{equ:qft_0C}
S(k)=\DST[s](k)= \sum_{n=1}^{\frac{N}{2}-1} s(n) \cdot \sin( \theta \cdot n \cdot k) \quad  k \in \{1,2,3,...,(\frac{N}{2}-1)\}
\end{equation}
where $N$ is the periodization of the transform applied to the signal (identical to the usual `length' term for $\DFT$) and $\theta$ is the angle pulse of fundamental frequency, defined as:
\begin{equation} \label{equ:qft_0D}
\theta=  \frac{2 \cdot \pi}{N}
\end{equation}
The $\DCT$ and $\DST$ transforms are defined in compliance with the definitions given in \cite{Guo_Sitton_qft_1998}, \cite{Pasquini_2013}.
We can call the $\DCT$ and $\DST$ transforms as $\DCT-0$ and $\DST-0$, to distinguish them from other $\DCT$ and $\DST$ types already defined in literature (however they are similar to $\DCT$-I and $\DST$-I respectively).
Let us observe that we can apply $\DFT$ transform both to real ($\RDFT$) and complex ($\CDFT$) valued signals.
With an abuse of notation, we will use the $\DFT$, $\DCT$, $\DST$ terms in case of pruned input and/or output too (when only a subset of $N$ values $s(n)$ are non-zero, or when only a subset of $N$ values $S(k)$ are required).

\emph{Sto\_n} (\emph{sto\_k}) describes the subset of $n$ ($k$) indices of a signal $s$ ($S$) that we store in memory.
For any signal used in this paper \emph{sto\_n}, \emph{sto\_k} have some relevant characteristics:
\begin{itemize}
\item 
$sto\_n$ coincides with the the group of only indices $n$ where $s(n)$ has not an a-priori known value.

\item
$sto\_k$ coincides with the required independent-value harmonics of a signal.

\end{itemize}
We define $ln$ ($lk$) as the storage size in temporal (frequency) domain, that represents the number of real value cells required to store the $sto\_n$, $sto\_k$ group. 

An univocal choise of transform ($\CDFT$, $\RDFT$, $\DCT$, $\DST$), $sto\_n$ group, $sto\_k$ group, represents a signal type, as listed in Tab.\ref{tab:notation}.

%%%%%%%%%%%%%%%%%%%%%%%%%%%%%%%%%%%%%%%%%%%%%%%

\subsection{Basic elaborations} \label{sec:basic_elaborations}

A basic elaboration is a way to process a signal inside an FFT algorithm that we do not need to split into simpler fundamental mathematical operations.
We use two kinds of basic elaborations: \emph{decompositions} and \emph{transformations}.
The former (latter) creates one (two) child(ren) signal(s) from the input signal.
Each basic elaboration is used in two phases: the \emph{forward phase} and the \emph{backward phase}. 
In the fomer we handle the temporal values and the known elements are the ones of mother input signal, the unknown elements are the ones of child(ren) signal(s). 
In the latter we handle the frequency-domain values and the known elements are the ones of child(ren) signal(s), while the unknown elements are the ones of input mother signal.

%%%%%%%%%%%%%%%%%%%%%%%%%%%%

\subsection{Elaboration diagrams} \label{sec:diagram}

An elaboration diagram is a concatenation of signals and arrows (i.e. Fig.\ref{fig:QFT_4_diagram}) that describes the sequence of descendent signals and basic elaborations respectively created and used inside a function in an FFT algorithm.

Each arrow corresponds to a basic elaboration. 
Beside each arrow we put the identifier of this basic elaboration (as ${M_4}$, ${H_k}$, ${D_k}$, etc.).
This graphical tool condenses any information of a function hiding (without loosing) any detail (that the reader can obtain using  sect.\ref{sec:transformations} and Tab.\ref{tab:elabora_relations_A},\ref{tab:elabora_relations_B},\ref{tab:DCT_odd_transformations},\ref{tab:DST_odd_transformations} for basic elaborations, and Tab.\ref{tab:notation} for signal types associated to any signal).
In sect.\ref{sec:improved_QFT} the reader can find an example on how to obtain the pseudo-code of a function, starting from its elaboration diagram.

%%%%%%%%%%%%%%%%%%%%%%%%%%%%%%%%%%%%%

\subsection{Notation for signals and signal types} \label{sec:notation}

We use a notation that creates a mnemonic link beetwen a signal (or signal type) name and its characteristics.

\subsubsection{Notation for signal types} \label{sec:notation_sub}

We associate a specific name to each signal type. 
Quoting from  \cite{Pasquini_2013}:
\begin{quote}
``

[\dots]
\begin{itemize}

\item
the main symbol is `s' (s=signal) $[\dots]$.

\item
the first subscript symbol identifies the applied transform: `$cx$'  (complex $\DFT$), `$re$' (real $\DFT$), `$dc$' ($\DCT$), `$ds$' ($\DST$).

\item
the second subscript symbol refers to $sto\_n$: `$o$' means generic odd, `$e$' or `$e_1$' are two different grouping of only even $n$ indices,  `$t$' or `$t_1$' (generic total) are two different grouping of both even and odd $n$ indices.

\item
the third subscript symbol refers to $sto\_k$: `$o$' (generic odd), `$e$' (generic even), `$t$' or `$t_1$' (generic total).

\end{itemize}
This notation highlights the parallelism in the elaboration used in the corresponding recursive functions, in the $\DCT$  context, and in the $\DST$ context, inside $[\dots]$ improved QFT algorithms. 
In this way, for many functions $[\dots]$ we can switch between signal types used in $\DCT$ context, to the ones used in $\DST$ context of the same algorithm, simply replacing the `$dc$' by the `$ds$' subscript (and keeping unchanged the remaining subscripts).
As a side effect of this notation, there is no univocal correspondence among a single subscript symbol, and a single feature of the signal (except for the 1st subscript), but only among a sequence of subscript symbols, and a signal type.
For example, the subscript `$e$' referring to $sto\_k$ identifies:
\begin{itemize}

\item
the group $sto\_k=\{0,2,\dots,(\frac{N}{2})\}$ if it is used in $s_{dc\_te}$ sequence of symbols.

\item
the group $sto\_k=\{2,4, \dots, (\frac{N}{2}-2)\}$ if it is used in $s_{ds\_te}$ sequence of symbols.

\end{itemize}
Notice that the exposed notation for signal types does not require to distinguish the `$t$' symbol (or any other symbol) depending on whether it refers to the grouping $sto\_n$, or it refers to the grouping  $sto\_k$ (for example using the $t_n$ in the first case, and the symbol $t_k$ in the second case), because we only need to consider the position of the symbol in the notation to see if it relates to $sto\_n$ or $sto\_k$. 
This choice has the advantage to make the name of each signal shorter.
Moreover this notation has the advantage that reading a signal type name we can immediately remember many characteristics of this signal.
For instance, reading the term $s_{ds\_to}$, we remember that it denotes the signal type to whom we apply the $\DST$, having both some even and odd residual time $n$ indices, and for which only some odd $k$ harmonics are required.
Tab.\ref{tab:notation} reports all and only the sequences of symbols (signal types) used in this paper.

''
\end{quote}
Moreover, for compactness reason, in Tab.\ref{tab:notation} we use the main symbol `$s$' (`$S$') to handle temporal (frequency-domain) values of a signal type.

\subsubsection{Notation for signals} \label{sec:notation_sub_2}

An FFT algorithm can create many different signals which share the same signal type.
We use a mnemonic notation for signals too.
In sect.\ref{sec:improved_QFT}, in Tab.\ref{tab:DCT_odd_transformations},\ref{tab:DST_odd_transformations} and in Fig.\ref{fig:QFT_4_diagram},\ref{fig:QFT_2_diagram},\ref{fig:QFT_5_diagram},\ref{fig:QFT_6_diagram},\ref{fig:QFT_2_tree}, we use the same notation for signals already used in \cite{Pasquini_2013}.

We quote again from \cite{Pasquini_2013}:
\begin{quote}

``

[\dots]
Each used signal is described by means of a notation that slightly modifies the notation used for the associated signal type, according to these rules:
\begin{itemize}
\item 
the 1st symbol is `$s$' for temporal signals, and `$S$' for frequency-domain signals.

\item
an optional subscript identifier (numbers and/or capital letters), can be inserted after the 1st `$s|S$' symbol, to distinguish the handled signal from other signals of the same type used in the same context.

\end{itemize}
For example $s_{dc\_tt}$ and $s_{A\_dc\_tt}$, $s_{3,1\_A\_dc\_tt}$ are three different temporal signals of the same type `$s_{dc\_tt}$', while $S_{ds\_ot}$ and $S_{A\_ds\_ot}$, $S_{4,7\_A\_ds\_ot}$ are three different frequency-domain signals of the same type `$s_{ds\_ot}$'.

''

\end{quote}
Moreover we add `$[N]$' at the end of a signal name to state that we apply a transform with periodization $N$ to it.
For example the signals $s_{dc\_oo}[N]$ and $s_{A\_dc\_oo}[\frac{N}{2}]$ created in $dct\_oo$ function in improved QFT (see Fig.\ref{fig:QFT_4_diagram}) are two different signals that share the same signal type, but with a different periodization.

Differently in order to concise the description of basic elaborations applied to many input signals (with different associated signal types), only in sect.\ref{sec:le_varianti_della_QFT} and in Tab.\ref{tab:elabora_relations_A},\ref{tab:elabora_relations_B} we use a not detailed notation (such as $s_{on}$, $s_{en}$, $S_{ek}$, $s_{A}$, $s_{B}$, etc.).

\begin{table}[tb]
\caption{Transform type, $sto\_n$ and $sto\_k$ indices, ln and lk parameters, associated to any signal type used in this paper.}
\label{tab:notation}
\centering
\scalebox{0.9}
{
\begin{tabular}{cccccc}
\toprule
signal type & transform type & sto\_n & sto\_k & ln & lk\\
\midrule
$s_{cx\_tt}$ & $\CDFT$ & $\{ 0,1,2,\dots,(N-1) \}$ & $ \{ 0,1,2,\dots,(N-1) \}$ & $2 \cdot N$ & $2 \cdot N$ \\
$s_{re\_tt}$ & $\RDFT$ & $\{ 0,1,2,\dots,(N-1) \}$ & $ \{ 0,1,2,\dots,(N-1) \}$  & $N$ & $N$ \\
$s_{dc\_tt}$ & $\DCT$ & $\{ 0,1,2,\dots,(\frac{N}{2}) \}$ & $ \{ 0,1,2,\dots,(\frac{N}{2}) \}$  & $\frac{N}{2}+1$ & $\frac{N}{2}+1$ \\
$s_{dc\_et}$ & $\DCT$ & $\{ 0,2,4,\dots,(\frac{N}{2}) \}$ & $\{ 0,1,2,\dots,(\frac{N}{4}) \}$  & $\frac{N}{4}+1$ & $\frac{N}{4}+1$ \\
$s_{dc\_ot}$ & $\DCT$ & $\{ 1,3,5,\dots,(\frac{N}{2}-1) \}$ & $ \{ 0,1,2,\dots,(\frac{N}{4}-1) \}$  & $\frac{N}{4}$ & $\frac{N}{4}$ \\
$s_{dc\_te}$ & $\DCT$ & $\{ 0,1,2,\dots,(\frac{N}{4}) \}$ & $ \{ 0,2,4,\dots,(\frac{N}{2}) \}$  & $\frac{N}{4}+1$ & $\frac{N}{4}+1$ \\
$s_{dc\_to}$ & $\DCT$ & $\{ 0,1,2,\dots,(\frac{N}{4}-1) \}$ & $ \{ 1,3,5,\dots,(\frac{N}{2}-1) \}$  & $\frac{N}{4}$ & $\frac{N}{4}$ \\
$s_{dc\_oe}$ & $\DCT$ & $ \{ 1,3,5,\dots,(\frac{N}{4}-1) \}$ & $ \{ 0,2,4,\dots,(\frac{N}{4}-2) \}$ & $\frac{N}{8}$ & $\frac{N}{8}$\\
$s_{dc\_eo}$ & $\DCT$ &  $\{ 0,2,4,\dots,(\frac{N}{4}-2) \}$ & $ \{ 1,3,5,\dots,(\frac{N}{4}-1) \}$ & $\frac{N}{8}$ & $\frac{N}{8}$\\
$s_{dc\_oo}$ & $\DCT$ & $\{ 1,3,5,\dots,(\frac{N}{4}-1) \}$ & $\{ 1,3,5,\dots,(\frac{N}{4}-1) \}$ & $\frac{N}{8}$ & $\frac{N}{8}$\\
$s_{ds\_tt}$ & $\DST$ & $\{ 1,2,3,\dots,(\frac{N}{2}-1) \}$ & $ \{ 1,2,3,\dots,(\frac{N}{2}-1) \}$ & $\frac{N}{2}-1$ & $\frac{N}{2}-1$\\
$s_{ds\_et}$ & $\DST$ & $ \{ 2,4,6,\dots,(\frac{N}{2}-2) \}$ & $ \{ 1,2,3,\dots,(\frac{N}{4}-1) \}$ & $\frac{N}{4}-1$ & $\frac{N}{4}-1$\\
$s_{ds\_te}$ & $\DST$ & $\{ 1,2,3,\dots,(\frac{N}{4}-1) \}$ & $ \{ 2,4,6,\dots,(\frac{N}{2}-1) \}$ & $\frac{N}{4}-1$ & $\frac{N}{4}-1$\\
$s_{ds\_to}$ & $\DST$ & $ \{ 1,2,3,\dots,(\frac{N}{4}) \}$ & $ \{ 1,3,5,\dots,(\frac{N}{2}-1) \}$ & $\frac{N}{4}$ & $\frac{N}{4}$\\
$s_{ds\_ot}$ & $\DST$ & $\{ 1,3,5,\dots,(\frac{N}{2}-1) \}$ & $\{ 1,2,3,\dots,(\frac{N}{4}) \}$ & $\frac{N}{4}$ & $\frac{N}{4}$\\
$s_{ds\_oe}$ & $\DST$ & $\{ 1,3,5,\dots,(\frac{N}{4}-1) \}$ & $ \{ 2,4,6,\dots,(\frac{N}{4}) \}$ & $\frac{N}{8}$ & $\frac{N}{8}$\\
$s_{ds\_eo}$ & $\DST$ &  $\{ 2,4,6,\dots,(\frac{N}{4}) \}$ &  $\{ 1,3,5,\dots,(\frac{N}{4}-1) \}$ & $\frac{N}{8}$ & $\frac{N}{8}$\\
$s_{ds\_oo}$ & $\DST$ &  $\{1,3,5,\dots,(\frac{N}{4}-1)\}$ & $ \{ 1,3,5,\dots,(\frac{N}{4}-1) \}$ & $\frac{N}{8}$ & $\frac{N}{8}$\\
\bottomrule
\end{tabular}
}
\end{table}

%%%%%%%%%%%%%%%%%%%%%%%%%%%%%%%%%

\begin{table}[tb]
\caption{a possible matching between the theoretical signal (i.e. $s_{cx\_tt}$) and the array of memory cells (i.e. $s_{cx\_tt\_arr}$ or $S_{cx\_tt\_arr}$), in an implementation where each signal is stored into a contiguous sequence of cell (array), indices $n$ as $k$ are stored in growing order, and the first cell of an array has index $p=1$.}
\label{tab:implementation}
\centering
\scalebox{0.7}
{
\begin{tabular}{ccc}
\toprule
signal type & matching temporal signal s - array & matching frequency-domain signal S - array \\
\midrule
$s_{cx\_tt}$ & $s_{cx\_tt}(n)=s_{cx\_tt\_arr}(n+1)$ & $S_{cx\_tt}(k)=S_{cx\_tt\_arr}(k+1)$ \\
$s_{re\_tt}$ & $s_{re\_tt}(n)=s_{re\_tt\_arr}(n+1)$ & $S_{re\_tt}(k)=S_{re\_tt\_arr}(k+1)$ \\
$s_{dc\_tt}$ & $s_{dc\_tt}(n)=s_{dc\_tt\_arr}(n+1)$ & $S_{dc\_tt}(k)=S_{dc\_tt\_arr}(k+1)$ \\
$s_{dc\_et}$ & $s_{dc\_et}(n)=s_{dc\_et\_arr}(\frac{n+2}{2})$ & $S_{dc\_et}(k)=S_{dc\_et\_arr}(k+1)$ \\
$s_{dc\_ot}$ & $s_{dc\_ot}(n)=s_{dc\_ot\_arr}(\frac{n+1}{2})$ & $S_{dc\_ot}(k)=S_{dc\_ot\_arr}(k+1)$ \\
$s_{dc\_te}$ & $s_{dc\_te}(n)=s_{dc\_te\_arr}(n+1)$ & $S_{dc\_te}(k)=S_{dc\_te\_arr}(\frac{k+2}{2})$ \\
$s_{dc\_to}$ & $s_{dc\_to}(n)=s_{dc\_to\_arr}(n+1)$ & $S_{dc\_to}(k)=S_{dc\_to\_arr}(\frac{k+1}{2})$ \\
$s_{dc\_oe}$ & $s_{dc\_oe}(n)=s_{dc\_oe\_arr}(\frac{n+1}{2})$ & $S_{dc\_oe}(k)=S_{dc\_oe\_arr}(\frac{k+2}{2})$ \\
$s_{dc\_eo}$ & $s_{dc\_eo}(n)=s_{dc\_eo\_arr}(\frac{n+2}{2})$ & $S_{dc\_eo}(k)=S_{dc\_eo\_arr}(\frac{k+1}{2})$ \\
$s_{dc\_oo}$ & $s_{dc\_oo}(n)=s_{dc\_oo\_arr}(\frac{n+1}{2})$ & $S_{dc\_oo}(k)=S_{dc\_oo\_arr}(\frac{k+1}{2})$ \\
$s_{ds\_tt}$ & $s_{ds\_tt}(n)=s_{ds\_tt\_arr}(n)$ & $S_{ds\_tt}(k)=S_{ds\_tt\_arr}(k)$ \\
$s_{ds\_et}$ & $s_{ds\_et}(n)=s_{ds\_et\_arr}(\frac{n}{2})$ & $S_{ds\_et}(k)=S_{ds\_et\_arr}(k)$ \\
$s_{ds\_ot}$ & $s_{ds\_ot}(n)=s_{ds\_ot\_arr}(\frac{n+1}{2})$ & $S_{ds\_ot}(k)=S_{ds\_ot\_arr}(k)$ \\
$s_{ds\_te}$ & $s_{ds\_te}(n)=s_{ds\_te\_arr}(n)$ & $S_{ds\_te}(k)=S_{ds\_te\_arr}(\frac{k}{2})$ \\
$s_{ds\_to}$ & $s_{ds\_to}(n)=s_{ds\_to\_arr}(n)$ & $S_{ds\_to}(k)=S_{ds\_to\_arr}(\frac{k+1}{2})$ \\
$s_{ds\_oe}$ & $s_{ds\_oe}(n)=s_{ds\_oe\_arr}(\frac{n+1}{2})$ & $S_{ds\_oe}(k)=S_{ds\_oe\_arr}(\frac{k}{2})$ \\
$s_{ds\_eo}$ & $s_{ds\_eo}(n)=s_{ds\_eo\_arr}(\frac{n}{2})$ & $S_{ds\_eo}(k)=S_{ds\_eo\_arr}(\frac{k+1}{2})$ \\
$s_{ds\_oo}$ & $s_{ds\_oo}(n)=s_{ds\_oo\_arr}(\frac{n+1}{2})$ & $S_{ds\_oo}(k)=S_{ds\_oo\_arr}(\frac{k+1}{2})$ \\
\bottomrule
\end{tabular}
}
\end{table}

%%%%%%%%%%%%%%%%%%%%%%%%%%%%%%%%%%%%%%%%%%%%%%%%%%%%%%%%

\section{Common basic elaborations used in developed algorithms} \label{sec:transformations}

The eight algorithms described in this paper share some common basic elaborations (decomposition or transformations).
Some of them are applied to a single signal type: the decomposition of $\CDFT$ into two $\RDFT$ and the decomposition of RDFT into the couple ($\DCT$, $\DST$). 
Conversely the others are applied to many signal types (and for this reason we describe mathematical details of these elaborations, in general case, without specifying which signal types are involved): the separation of even harmonics from odd ones, the separation of even time indices from odd ones, the even harmonics halving and the even time indices halving. 

We describe here, as an example, two basic elaborations in natural language: the decomposition of $\CDFT$ into two $\RDFT$, and the decomposition of $\RDFT$ into the couple 
$\DCT$ and $\DST$. 
The remaining elaborations (that are applied to many mother signal types in this paper) are described in Tab.\ref{tab:elabora_relations_A},\ref{tab:elabora_relations_B},\ref{tab:DCT_odd_transformations},\ref{tab:DST_odd_transformations}.
They briefly report both temporal and frequency-domain relations involved by these basic elaborations, the signal types received as input, and the ones created as output.
The reader can find a more detailed description (in natural language) of many of these tabulated basic elaborations in \cite{Pasquini_2013}.

\subsection{The decomposition $D_c$ of $\CDFT$ into two $\RDFT$}

We quote from \cite{Pasquini_2013}:

``

[\dots] The input signal of the decomposition of $\CDFT$ into two $\RDFT$, is only of type $s_{cx\_tt}$. Let's call $N$ its length (equal to periodization).
This elaboration decomposes the $\CDFT$ calculation into two $\RDFT$ transforms, relative to children output signals $s_{1\_re\_tt}$ and $s_{2\_re\_tt}$, both of length $N$ (equal to periodization). 
\begin{align} 
& s_{1\_re\_ tt}(n)= \Re[s_{cx\_ tt}(n)] \quad n \in \{0,1,2,\dots,(N-1) \} \label{equ:qft_15} \\
& s_{2\_re\_ tt}(n)= \Im[s_{cx\_ tt}(n)] \quad n \in \{0,1,2,\dots,(N-1) \} \label{equ:qft_16}
\end{align}
We can prove that eq.(\ref{equ:qft_15}),(\ref{equ:qft_16}) correspond to the following frequency-domain relationships (backward phase):
\begin{equation}
\begin{split} \label{equ:qft_17}   
\Re \{\CDFT[ s_{cx\_tt}]\}(k) & = \Re \{\RDFT[s_{1\_re\_ tt}]\}(k) - \Im \{\RDFT[s_{2\_re\_tt}] \}(k) \\
& \quad k \in \{1,2,\dots,(\frac{N}{2}-1)\} 
\end{split} \\
\end{equation}
\begin{equation}
\begin{split} \label{equ:qft_18}
\Re \{\CDFT[ s_{cx\_tt}]\}(N-k) & = \Re \{\RDFT[s_{1\_re\_tt}]\}(k) + \Im \{\RDFT[s_{2\_re\_tt}] \}(k)  \\
& \quad  k \in  \{1,2,\dots,(\frac{N}{2}-1)\} 
\end{split} \\
\end{equation}
\begin{equation}
\Re \{\CDFT[ s_{cx\_tt}]\}(k)= \Re \{\RDFT[s_{1\_re\_tt}]\}(k) \quad k \in \{ 0, (\frac{N}{2}) \} \label{equ:qft_19}
\end{equation}
\begin{equation}  
\begin{split} \label{equ:qft_19A}
 \Im \{\CDFT[ s_{cx\_tt}]\}(k) & = \Im \{\RDFT[s_{1\_re\_tt}]\}(k) - \Re\{\RDFT[s_{2\_re\_tt}] \}(k) \\
& \quad k \in \{1,2,\dots,(\frac{N}{2}-1)\}  
\end{split} \\
\end{equation}
\begin{equation}
\begin{split} \label{equ:qft_20}
 \Im \{\CDFT[ s_{cx\_tt}]\}(N-k) & = -\Im \{\RDFT[s_{1\_re\_tt}]\}(k) + \Re \{\RDFT[s_{2\_re\_tt}] \}(k) \\
& \quad k \in  \{1,2,\dots,(\frac{N}{2}-1)\}
\end{split} \\
\end{equation}
\begin{equation}
  \Im \{\CDFT[ s_{cx\_tt}]\}(k) = \Re \{\RDFT[s_{2\_re\_tt}]\}(k) \quad k \in \{ 0, (\frac{N}{2}) \} \label{equ:qft_21}
\end{equation}

''

\subsection{The decomposition $D_r$ of $\RDFT$ into two $\DCT$ and $\DST$}

We quote from \cite{Pasquini_2013}:

``

[\dots] The input signal of the decomposition of $\RDFT$ into $\DCT$ and $\DST$, is only of type $s_{re\_tt}$ in this paper. Let's call $N$ its length (equal to periodization).
This elaboration decomposes the $\RDFT$ calculation into the calculation of a $\DCT$ (applied to the child signal $s_{dc\_tt}$ of periodization equal to $N$ [\dots]) and a $\DST$ (applied to the child signal $s_{ds\_tt}$ of periodization equal to $N$ [\dots]). 
We can prove that the following time domain equations hold:
\begin{equation} \label{equ:qft_22}
s_{dc\_tt}(n)= \left\{ \begin{array}{ll}
s_{re\_tt}(n) + s_{re\_tt}(N-n) \quad & n \in \{1,2,3,\dots,(\frac{N}{2}-1) \} \\
s_{re\_tt}(n) \quad & n \in \{0, (\frac{N}{2}) \} \\
0 \quad & \text{otherwise}  \\
\end{array} \right.
\end{equation}
\begin{equation} \label{equ:qft_23}
s_{ds\_tt}(n)= \left\{ \begin{array}{ll}
s_{re\_tt}(n) - s_{re\_tt}(N-n) \quad & n \in \{1,2,3,\dots,(\frac{N}{2}-1) \} \\
0 \quad & \text{otherwise}  \\
\end{array} \right.
\end{equation}
corresponding to the following frequency-domain relationships (backward phase):
\begin{align}
& \Re \{\RDFT[s_{re\_tt}]\}(k) = \DCT[s_{dc\_tt}](k) \quad & k \in \{ 0,1,2,\dots,(\frac{N}{2}) \} \label{equ:qft_24} \\
& \Im \{\RDFT[s_{re\_tt}]\}(k) = - \DST[s_{ds\_tt}](k) \quad & k \in \{ 1,2,3,\dots,(\frac{N}{2}-1) \} \label{equ:qft_25}
\end{align}

''

\begin{table}[tb]
\caption{basic elaborations: temporal and frequency-domain relations and involved signal types, for even harmonics halving and for even time indices halving}
\label{tab:elabora_relations_A}
\centering
\scalebox{0.75}
{
\begin{tabular}{c}
\toprule

$H_K$: Even Harmonics Halving  \\

$s_{t_k}(n) = \left\{ \begin{array}{ll} s_{e_k}(n) \quad &  n \in sto\_n(s_{e_k}) \\ 0 \quad & \text{otherwise}  \\ \end{array} \right.$  \\

$S_{e_k}(k=2 \cdot k_A) = S_{t_k}(k_A) \quad k \in sto\_k(s_{e_k})$ \\
\\
if $s_{e_k}=s_{dc\_te}$ then $s_{t_k}=s_{dc\_tt}$.\\
if $s_{e_k}=s_{dc\_oe}$ then $s_{t_k}=s_{dc\_ot}$.\\
if $s_{e_k}=s_{ds\_te}$ then $s_{t_k}=s_{ds\_tt}$.\\
if $s_{e_k}=s_{ds\_oe}$ then $s_{t_k}=s_{ds\_ot}$.\\

\\

$H_n$: Even Time-indices Halving \\

$s_{t_n}(n_A) = \left\{ \begin{array}{ll} s_{e_n}(n=2 \cdot n_A) \quad &  n_A \quad \in sto\_n(s_{t_n}) \\ 0 \quad & \text{otherwise}  \\ \end{array} \right.$ \\ $\DCT[s_{e_n}](k) = \DCT[s_{t_n}](k) \quad k \in sto\_k(s_{e_n})$ \\

$\DST[s_{e_n}](k) = \DST[s_{t_n}](k) \quad k \in sto\_k(s_{e_n})$ \\
\\
if $s_{e_n}=s_{dc\_et}$ then $s_{t_n}=s_{dc\_tt}$. \\
if $s_{e_n}=s_{dc\_eo}$ then $s_{t_n}=s_{dc\_to}$. \\
if $s_{e_n}=s_{ds\_et}$ then $s_{t_n}=s_{ds\_tt}$. \\
if $s_{e_n}=s_{ds\_eo}$ then $s_{t_n}=s_{ds\_to}$. \\

\\

\bottomrule
\end{tabular}
} 
\end{table}

\begin{table}[tb]
\caption{basic elaborations: temporal and frequency-domain relations and involved signal types, for separation of even harmonics from odd ones and for separation of even time indices from odd ones}
\label{tab:elabora_relations_B}
\centering
\scalebox{0.75}
{
\begin{tabular}{c}
\toprule

$D_K$: Separation of even harmonics from odd ones (in $\DCT$ context) \\

$s_{e_k}(n) = \left\{ \begin{array}{ll}  s_{t_k}(n) + s_{t_k}(\frac{N}{2}-n) \quad & n \in sto\_n(s_{e_k})/\{n=\frac{N}{4}\}  \\ s_{t_k}(n) \quad &  \{n=\frac{N}{4}\} \cap sto\_n(s_{e_k}) \\ 0 \quad & \text{otherwise}  \\ \end{array} \right.$ \\

$S_{t_k}(n) = \left\{ \begin{array}{ll} S_{e_k}(k) \quad & k \quad \text{even} \in sto\_k(s_{e_k}) \\ s_{o_k}(k) \quad & k \quad \text{odd} \in sto\_k(s_{o_k}) \\ \end{array} \right.$  \\

$s_{o_k}(n) = \left\{ \begin{array}{ll} s_{t_k}(n) - s_{t_k}(\frac{N}{2}-n) \quad & n \in sto\_n(s_{o_k})  \\ 0    \quad & \text{otherwise}  \\ \end{array} \right.$ \\

\\

if $s_{t_k}=s_{dc\_tt}$ then $s_{o_k}=s_{dc\_to}$ and $s_{e_k}=s_{dc\_te}$. \\
if $s_{t_k}=s_{dc\_ot}$ then $s_{o_k}=s_{dc\_oo}$ and $s_{e_k}=s_{dc \_oe}$. \\
if $s_{t_k}=s_{ds\_tt}$ then $s_{o_k}=s_{ds\_to}$ and $s_{e_k}=s_{ds\_te}$. \\
if $s_{t_k}=s_{ds\_ot}$ then $s_{o_k}=s_{ds\_oo}$ and $s_{e_k}=s_{ds\_oe}$. \\

\\

$D_K$: Separation of even harmonics from odd ones (in $\DST$ context) \\

$s_{e_k}(n)  = \left\{ \begin{array}{ll} s_{t_k}(n) - s_{t_k}(\frac{N}{2}-n) \quad & n \in sto\_n(s_{e_k}) \\ 0 \quad & \text{otherwise}  \\ \end{array} \right.$ \\ $S_{t_k}(n) = \left\{ \begin{array}{ll} S_{e_k}(k) \quad & k \quad \text{even} \in sto\_k(s_{e_k}) \\ S_{o_k}(k) \quad & k \quad \text{odd} \in sto\_k(s_{o_k}) \\ \end{array} \right.$  \\

 $s_{o_k}(n) = \left\{ \begin{array}{ll} s_{t_k}(n) + s_{t_k}(\frac{N}{2}-n) \quad & n \in sto\_n(s_{o_k})/\{n=\frac{N}{4}\} \\ s_{t_k}(n) \quad & \{n=\frac{N}{4}\}  \cap sto\_n(s_{e_k}) \\ 0 \quad & \text{otherwise}  \\ \end{array} \right.$ \\

\\

if $s_{t_k}=s_{dc\_tt}$ then $s_{o_k}=s_{dc\_to}$ and $s_{e_k}=s_{dc\_te}$. \\
if $s_{t_k}=s_{dc\_ot}$ then $s_{o_k}=s_{dc\_oo}$ and $s_{e_k}=s_{dc \_oe}$. \\
if $s_{t_k}=s_{ds\_tt}$ then $s_{o_k}=s_{ds\_to}$ and $s_{e_k}=s_{ds\_te}$. \\
if $s_{t_k}=s_{ds\_ot}$ then $s_{o_k}=s_{ds\_oo}$ and $s_{e_k}=s_{ds\_oe}$. \\

\\

$D_n$: Separation of even time indices from odd ones (in $\DCT$ context) \\

$ s_{e_n}(n) = \left\{ \begin{array}{ll} s_{t_n}(n) \quad &  n \quad \text{even} \in sto\_n(s_{t_n}) \\ 0 \quad & \text{otherwise}  \\ \end{array} \right.$ \\ 

$\DCT[s_{t_n}](k) = \DCT[s_{e_n}](k) + \DCT[s_{o_n}](k) \quad  k \in sto\_k(s_{o_n})$ \\

$ s_{o_n}(n) = \left\{ \begin{array}{ll} s_{t_n}(n) \quad &  n \quad \text{odd} \in sto\_n(s_{t_n}) \\ 0 \quad & \text{otherwise}  \\ \end{array} \right.$ \\ 

$\DCT[s_{t_n}](\frac{N}{2}-k) = \DCT[s_{e_n}](k) - \DCT[s_{o_n}](k) \quad  k \in sto\_k(s_{o_n})$ \\

$\DCT[s_{t_n}](k) = \DCT[s_{e_n}](k)  \quad \{k=\frac{N}{4}\}  \cap sto\_k(s_{e_n}) $ \\

\\

if $s_{t_k}=s_{dc\_tt}$ then $s_{o_k}=s_{dc\_to}$ and $s_{e_k}=s_{dc\_te}$. \\
if $s_{t_k}=s_{dc\_ot}$ then $s_{o_k}=s_{dc\_oo}$ and $s_{e_k}=s_{dc \_oe}$. \\
if $s_{t_k}=s_{ds\_tt}$ then $s_{o_k}=s_{ds\_to}$ and $s_{e_k}=s_{ds\_te}$. \\
if $s_{t_k}=s_{ds\_ot}$ then $s_{o_k}=s_{ds\_oo}$ and $s_{e_k}=s_{ds\_oe}$. \\

\\

$D_n$: Separation of even time indices from odd ones (in $\DST$ context) \\

 $ s_{e_n}(n) = \left\{ \begin{array}{ll} s_{t_n}(n) \quad &  n \quad \text{even} \in sto\_n(s_{t_n}) \\ 0 \quad & \text{otherwise}  \\ \end{array} \right.$ \\ 
 
 $\DST[s_{t_n}](k) = \DST[s_{o_n}](k) + \DST[s_{e_n}](k)  \quad  k \in sto\_k(s_{e_n})$ \\

 $ s_{o_n}(n) = \left\{ \begin{array}{ll} s_{t_n}(n) \quad &  n \quad \text{odd} \in sto\_n(s_{t_n}) \\ 0 \quad & \text{otherwise}  \\ \end{array} \right.$ \\

 $\DST[s_{t_n}](\frac{N}{2}-k) = \DST[s_{o_n}](k) - \DST[s_{e_n}](k) \quad  k \in sto\_k(s_{e_n})$ \\

 $\DST[s_{t_n}](k) = \DST[s_{o_n}](k)  \quad  \{k=\frac{N}{4}\}  \cap sto\_k(s_{o_n})$ \\

\\

if $s_{t_k}=s_{dc\_tt}$ then $s_{o_k}=s_{dc\_to}$ and $s_{e_k}=s_{dc\_te}$. \\
if $s_{t_k}=s_{dc\_ot}$ then $s_{o_k}=s_{dc\_oo}$ and $s_{e_k}=s_{dc \_oe}$. \\
if $s_{t_k}=s_{ds\_tt}$ then $s_{o_k}=s_{ds\_to}$ and $s_{e_k}=s_{ds\_te}$. \\
if $s_{t_k}=s_{ds\_ot}$ then $s_{o_k}=s_{ds\_oo}$ and $s_{e_k}=s_{ds\_oe}$. \\

\\

\bottomrule
\end{tabular}
} 
\end{table}

%%%%%%%%%%%%%%%%%%%%%%%%%%%%%%%%%%%%%%%%%%%%%%%%%%%%%%5

\section{The improved QFT algorithm} \label{sec:improved_QFT}

\begin{figure}[htb]
  \centering
  \includegraphics[width=0.9 \textwidth]{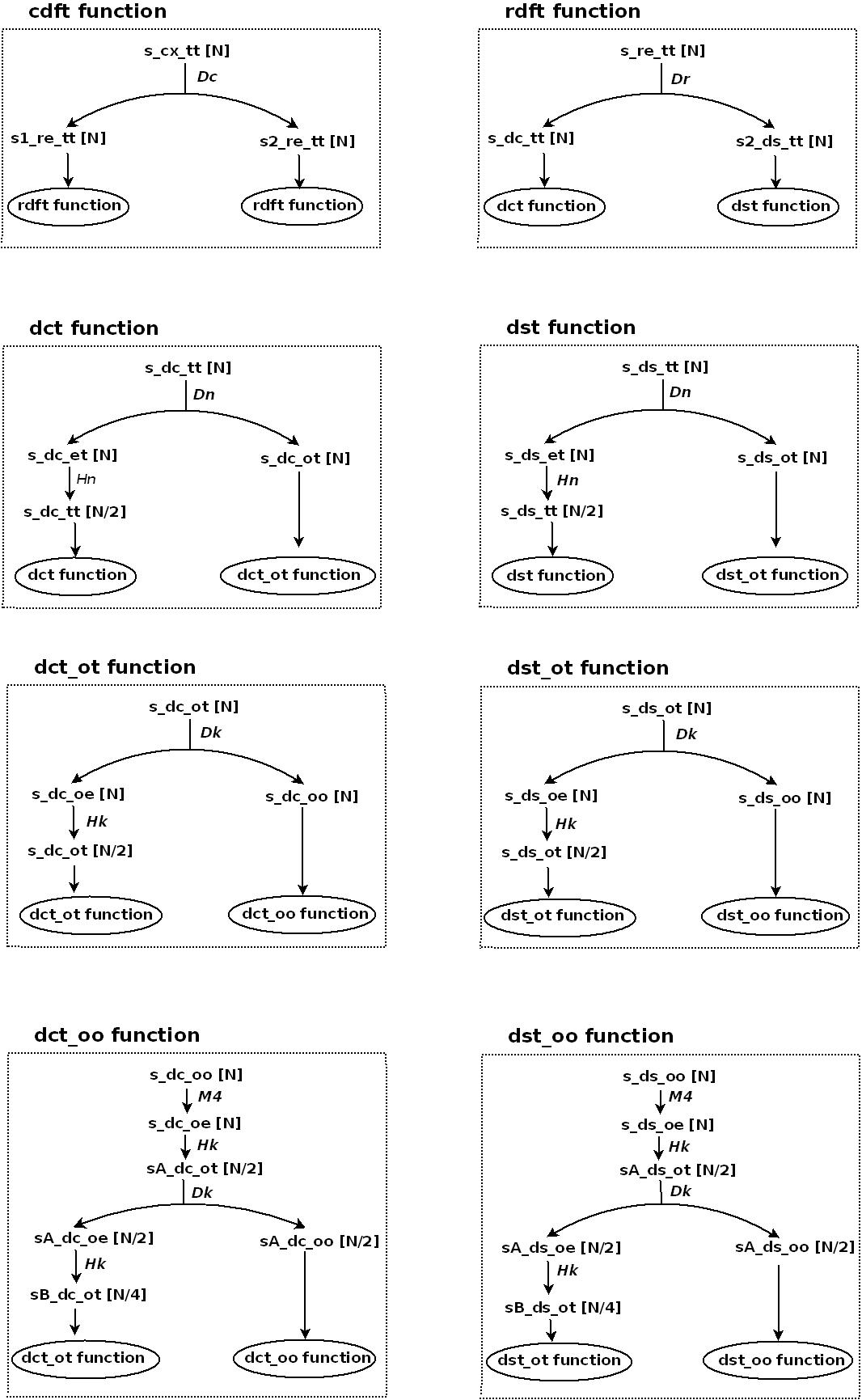}
  \caption{The elaboration diagrams of functions used in improved QFT (the 4th QFT variant)}
  \label{fig:QFT_4_diagram}
\end{figure}

The improved QFT algorithm is a real-factor algorithm which improves \cite{Pasquini_2013} the characteristics of classical QFT, obtaining qualities similar to split-radix 3add/3mul.
It can be described in terms of eight functions calling each other (if it is finalized to the computation of the $\CDFT$): $cdft$, $rdft$, $dct$, $dst$, $dct\_ot$, $dst\_ot$, $dct\_oo$ and $dst\_oo$.  
Each function decomposes the input signal into two output signals for any $N$, except in special cases ($N=8$ for $dct\_oo$ and $dst\_oo$, $N=4$ for $dst$, $dct\_ot$ and $dst\_ot$, $N=2$ for $dct$, $cdft$ and $rdft$), where we just apply the direct definition of the transform to the input signal.
We describe the improved QFT in a simpler, more compact manner with respect to  \cite{Pasquini_2013}, using the elaboration diagrams (defined in sect.\ref{sec:diagram}) shown in Fig.\ref{fig:QFT_4_diagram}.

The procedure that lets us to obtain the pseudo-code of a function, starting from its elaboration diagram, is made of three steps of back-abstraction.
The first step converts the basic diagram into a sequence of basic elaborations, described in an abstract way.
In this step we use the notation $E^{T}$ ($E^{F}$) to describe the forward (backward) phase of a basic elaboration $E$, where we handle the temporal (frequency-domain) elements. 
For example here is the abstract description (using basic elaboration identifiers ${M_4}$, ${H_k}$, ${D_k}$) of the function $dct\_oo$ of improved QFT.

\bigskip

\rule{\textwidth}{0.2mm}

function $dct\_oo$ used in IMPROVED QFT ALGORITHM  (abstract description)

\rule{\textwidth}{0.2mm}

\begin{align*}
& function  \quad prototype:  \quad \DCT[s_{dc\_oo}][N] \leftarrow dct\_oo(s_{dc\_oo}[N]); \\
& if \quad N>8\quad then \\
& \quad N_A=\frac{N}{2}; \quad N_B=\frac{N}{4}; \\
& \quad s_{dc\_oe}[N] \leftarrow M_4^{T}(s_{dc\_oo}[N]); \\
& \quad s_{A\_dc\_ot}[N_A] \leftarrow H_k^{T}(s_{dc\_oe}[N]); \\
& \quad [s_{A\_dc\_oe}[N_A], s_{A\_dc\_oo}[N_A]] \leftarrow D_k^{T}(s_{A\_dc\_ot}[N_A]); \\
& \quad s_{B\_dc\_ot}[N_B] \leftarrow H_k^{T}(s_{B\_dc\_oe}[N_A]); \\
& \quad S_{B\_dc\_ot}[N_B] \leftarrow dct\_ot(s_{B\_dc\_ot}[N_B]); \\
& \quad S_{A\_dc\_oo}[N_A] \leftarrow dct\_oo(s_{A\_dc\_oo}[N_A]); \\
& \quad S_{B\_dc\_oe}[N_A] \leftarrow H_k^{F}(S_{B\_dc\_ot}[N_B]); \\
& \quad S_{A\_dc\_ot}[N_A] \leftarrow D_k^{F}[(S_{A\_dc\_oe}[N_A]), S_{A\_dc\_oo}[N_A]); \\
& \quad S_{dc\_oe}[N] \leftarrow H_k^{F}(S_{A\_dc\_ot}[N_A]); \\
& \quad S_{dc\_oo}[N] \leftarrow M_4^{F}(S_{\_dc\_oe}[N]); \\
& else \quad (\text{direct definition of $\DCT$ is applied:}) \\
& \quad S_{dc\_oo}[N] \leftarrow \DCT(s_{\_dc\_oe}[N]); \\
& end \quad if;
\end{align*}

\rule{\textwidth}{0.2mm}

As we can see, we just have to follow the arrows in the elaboration diagram in Fig.\ref{fig:QFT_4_diagram} (in $dct\_oo$ function case), from top to down, to handle the temporal signals, and conversely, from bottom to up, when we handle the frequency-domain signals.

The 2nd step of the procedure consists in substituting each basic elaboration identifier with its mathematical details, that we can find in sect.\ref{sec:transformations} or in Tab.\ref{tab:elabora_relations_A},\ref{tab:elabora_relations_B},\ref{tab:DCT_odd_transformations},\ref{tab:DST_odd_transformations}.
For example we  change the abstract instruction $s_{dc\_oe}[N] \leftarrow M_4^{T}(s_{dc\_oo}[N])$ with the temporal eq. associated to $M_4$ elaboration, shown in Tab.\ref{tab:DCT_odd_transformations}:
\begin{equation*}
s_{dc \_oe}(n) =  s_{dc\_oo}(n) \cdot \frac{1}{2 \cdot \cos(\theta \cdot n)} \quad n \in \{ 1,3,5,\dots, (\frac{N}{4}-1) \}
\end{equation*}
Analogously, we change the the abstract instruction $S_{dc\_oo}[N] \leftarrow M_4^{F}(S_{dc\_oe}[N])$ with the frequency-domain eq. associated to $M_4$ elaboration, shown in Tab.\ref{tab:DCT_odd_transformations}:
\begin{align*}
& \DCT[s_{dc\_oo}](k) = \DCT[s_{dc \_oe}](k-1)+ \DCT[s_{dc\_oo}](k+1) \\
& \qquad k \in \{ 1,3,5, \dots, (\frac{N}{4}-3) \} \\
& \DCT[s_{dc\_oo}](k=\frac{N}{4}-1) = \DCT[s_{dc\_oe}](k=\frac{N}{4}-2)
\end{align*}
The 3rd step consists in substituting each signal with its associated array (that stores the signal in memory), according to Tab.\ref{tab:implementation}, or to an analogous table depending on the implementation of the algorithm.
The pseudo-codes of remaining functions of improved QFT (4th variant), and of other variants of QFT, can be obtained in an analogous manner.

% \end{comment}

%%%%%%%%%%%%%%%%%%%%%%%%%%%%%%%%%%%%%%%%%%%%

\section{Basic ideas behind the 8 AM-QFT variants} \label{sec:le_varianti_della_QFT}

In improved QFT we convert  (difficult to handle) odd indices signal types $s_{dc\_oo}$ and $s_{ds\_oo}$ into (much more easy to handle) even-indices signals types, multiplying them by secant function in time domain.
In this QFT context, we can pursue the same goal applying many other kind of conversion to $s_{dc\_oo}$ and $s_{ds\_oo}$ signal types, keeping unchanged the general structure of the algorithm, and quite maintaining the same good qualities of improved QFT algorithm.

These different ways to convert odd indices signals, into even indices signals, can be obtained from a new starting idea: the amplitude modulation Double SideBand - Suppressed Carrier (AM DSB-SC), between the modulating signal $s_A$ and an opportune sinusoidal oscillation, whose frequency is equal to the fundamental harmonic of modulating signal $s_A$, to obtain the modulated signal $s_B$.
This processing creates a correspondence between odd harmonics of $s_A$, and even harmonics of $s_B$, and viceversa, and for these reasons it can be applied to convert odd indices signal, into even indices signal:
\begin{equation} \label{equ:qft_34}
s_{B}(n) =  s_{A}(n) \cdot \cos(\theta \cdot n) \quad n \in sto\_n(s_A)
\end{equation}
\begin{equation}
\begin{split} \label{equ:qft_35A}
\DCT[s_{B}](k) & =  \frac{1}{2} \cdot [\DCT[s_{A}](k-1) + \DCT[s_{A}](k+1)]  \\
& \quad k \in sto\_k(s_B)  
\end{split} \\
\end{equation}
\begin{equation}
\begin{split} \label{equ:qft_35B}
\DST[s_{B}](k) & =  \frac{1}{2} \cdot [\DST[s_{A}](k-1) + \DST[s_{A}](k+1)] \\
& \quad k \in sto\_k(s_B) 
\end{split} \\
\end{equation}
In order to avoid the required divisions by two in frequency-domain equations, it is more convenient to modify eq.(\ref{equ:qft_34}) by coupling the 2 factor with the trigonometric function, so that:
\begin{align}
& s_{B}(n) =  s_{A}(n) \cdot 2 \cdot \cos(\theta \cdot n) \quad & n \in sto\_n(s_A) \label{equ:qft_36} \\
& \DCT[s_{B}](k) =  \DCT[s_{A}](k-1) + \DCT[s_{A}](k+1) \quad & k \in sto\_k(s_B) \label{equ:qft_37} \\
& \DST[s_{B}](k) =  \DST[s_{A}](k-1) + \DST[s_{A}](k+1) \quad & k \in sto\_k(s_B) \label{equ:qft_38}
\end{align}
The main advantage of this choice is that, in not `on the fly' algorithm implementation, the calculation of the product `$2 \cdot \cos(\theta \cdot n)$' can be performed a-priori and the constants `$2 \cdot \cos(\theta \cdot n)$', instead of `$ \cos(\theta \cdot n)$' can be memorized.

The idea of using the AM DSB-SC transformation has already appeared in \cite{Cho_Themes_1978}, but used in $\CDFT$ (instead of $\DCT$, $\DST$) context, and obtaining an higher computational cost compared to the one of this class of AM-QFT algorithms.

\subsection{The idea behind the 1st AM-QFT variant}

Let $s_{e_k}$ be a signal of whom we need to store (and to compute) frequecy-domain signal values only in even harmonics, and let $s_{o_k}$ be a signal of whom we need to store (and to compute) frequecy-domain signal values only in odd harmonics. 
If we denote $s_B=s_{e_k}$, and $s_A=s_{o_k}$, then eq.(\ref{equ:qft_36}),(\ref{equ:qft_37}),(\ref{equ:qft_38}) become:
\begin{align}
& s_{e_k}(n) =  s_{o_k}(n) \cdot 2 \cdot  \cos(\theta \cdot n) \quad & n \in sto\_n(s_{o_k}) \label{equ:qft_39} \\
& \DCT[s_{e_k}](k) =  \DCT[s_{o_k}](k-1) + \DCT[s_{o_k}](k+1) \quad & k \in sto\_k(s_{e_k}) \label{equ:qft_40} \\
& \DST[s_{e_k}](k) =  \DST[s_{o_k}](k-1) + \DST[s_{o_k}](k+1) \quad & k \in to\_k(s_{e_k}) \label{equ:qft_41A}
\end{align}
If the mother signal is $s_{o_k}=s_{dc\_oo}$ then we easily derive that the child signal is $s_{e_k}=s_{dc\_oe}$ by using (\ref{equ:qft_39}),(\ref{equ:qft_40}) and Tab.\ref{tab:notation}.
If we pose $k=0$ in (\ref{equ:qft_40}) then we have a particular case which requires to extend the $\DCT$ definition to the case $k=-1$, employing the same eq.(\ref{equ:qft_0B}).
In the backward phase, re-elaborating eq.(\ref{equ:qft_40}) we derive the unknown frequency-domain components $S_{o_k}$, starting from the known ones $S_{e_k}$ and using the previous particular case too. 
Thus, in this 1st variant, the transformation of odd $sto\_k$ indices mother signal $s_{dc\_oo}$, into even $sto\_k$ indices child signal $s_{dc\_oe}$, occurs by means of relations of Tab.\ref{tab:DCT_odd_transformations} in $M_1$ case.
Conversely, if the mother signal is $s_{o_k}=s_{ds \_oo}$ then, by using eq.(\ref{equ:qft_39}),(\ref{equ:qft_41A}) and Tab.\ref{tab:notation}, we derive that the child signal
is $s_{e_k}=s_{ds\_oe}$.
If we pose $k=(\frac{N}{4})$ in (\ref{equ:qft_41A}) then we have a particular case.
In the backward phase, re-elaborating eq.(\ref{equ:qft_41A}) we derive the unknown frequency-domain components $S_{o_k}$, starting from the known ones $S_{e_k}$ and from the $k=\frac{N}{4}$ particular case. 
Thus, in this 1st variant, the transformation of odd $sto\_k$ mother signal $s_{ds\_oo}$ into the even $sto\_k$ child signal $s_{ds\_oe}$ occurs by means of relations of Tab.\ref{tab:DST_odd_transformations}, in $M_1$ case.

\subsection{The idea behind the 4th AM-QFT variant}

In order to simplify the exposition, we prefer to anticipate the 4th variant case, which coincides with the improved QFT \cite{Pasquini_2013}.
The idea is similar to the the 1st variant case, the only difference being that we pose $s_A=s_{e_k}$ and $s_B=s_{o_k}$ in eq.(\ref{equ:qft_36}),(\ref{equ:qft_37}),(\ref{equ:qft_38}). Re-elaborating eq.(\ref{equ:qft_36}) we obtain:
\begin{align}
& s_{e_k}(n) = s_{o_k}(n) \cdot \frac{1}{2 \cdot \cos(\theta \cdot n)}  \quad & n \in sto\_n(s_{o_k}) \label{equ:qft_46} \\
& \DCT[s_{o_k}](k) =  \DCT[s_{e_k}](k-1) + \DCT[s_{e_k}](k+1) \quad & k \in sto\_k(s_{o_k}) \label{equ:qft_47} \\
& \DST[s_{o_k}](k) =  \DST[s_{e_k}](k-1) + \DST[s_{e_k}](k+1) \quad & k \in sto\_k(s_{o_k}) \label{equ:qft_48A}
\end{align}
If we pose $s_{o_k}=s_{dc\_oo}$ in (\ref{equ:qft_46}),(\ref{equ:qft_47}), or $s_{o_k}=s_{ds\_oo}$ in  (\ref{equ:qft_46}),(\ref{equ:qft_48A}), then we obtain the 4th variant, that creates the output signal types and relations described in Tab.\ref{tab:DCT_odd_transformations} and Tab.\ref{tab:DST_odd_transformations} respectively, in $M_4$ case. 
Moreover let us observe that eq.(\ref{equ:qft_46}),(\ref{equ:qft_47}) are used in classical QFT \cite{Pasquini_2013} too (if applied to different signal types with respect to the 4th variant).
It follows that both classical and improved QFT share the re-elaborated AM DSB-SC modulation idea, with this class of algorithms. 

%Si osservi come, nelle implementazioni `non al volo' dell'algoritmo, il calcolo della quantità `$\frac{1}{2 \cdot \cos(\theta \cdot n)}$', può essere svolto a priori, e dunque si possono archiviare in memoria le costanti del tipo `$\frac{1}{2 \cdot \cos(\theta \cdot n)}$'.

\subsection{The idea behind the 2nd AM-QFT variant}

Using duality we can transform the odd $sto\_n$ indices mother signal, into the even $sto\_n$ indices child signal, instead of transforming the odd $sto\_k$ indices mother signal, into the even  $sto\_k$ indices child signal of the previous cases. 
Transforming by duality eq.(\ref{equ:qft_39}) we derive:
\begin{equation}  \label{equ:qft_53A}
S_{e_n}(k) = S_{o_n}(k) \cdot 2 \cdot \cos(\theta \cdot k)  \quad k \in sto\_k(s_{o_n}) 
\end{equation}
Observing that in the frequency-domain we proceed backward, and therefore we derive the frequency-domain components $S_{o_n}(k)$ from the $S_{e_n}(k)$ ones, eq.(\ref{equ:qft_53A}) is re-elaborated as follows:
\begin{equation}  \label{equ:qft_53B}
S_{o_n}(k) = S_{e_n}(k) \cdot \frac{1}{2 \cdot \cos(\theta \cdot k)}  \quad  k \in sto\_k(s_{o_n}) 
\end{equation}
Applying eq.(\ref{equ:qft_53B}) to mother signal type $s_{o_n}=s_{dc\_oo}$ ($s_{o_n}=s_{ds\_oo}$) we obtain the output signal type, and the relations, described in Tab.\ref{tab:DCT_odd_transformations} (Tab.\ref{tab:DST_odd_transformations}) in $M_2$ case, that constitute the 2nd AM-QFT variant.

\subsection{The idea behind the 3rd AM-QFT variant}

Applying duality to eq.(\ref{equ:qft_46}), and elaborating it in order to derive the frequency-domain components $S_{on}(k)$ from the $S_{en}(k)$, we obtain:
\begin{equation}  \label{equ:qft_59}
S_{o_n}(k) = S_{e_n}(k) \cdot 2 \cdot \cos(\theta \cdot k)  \quad  k \in sto\_k(s_{o_n}) 
\end{equation}
Applying eq.(\ref{equ:qft_59}) to mother signal type $s_{o_n}=s_{dc\_oo}$ ($s_{o_n}=s_{ds\_oo}$) we obtain the output signal type, and the relations, described in Tab.\ref{tab:DCT_odd_transformations} (Tab.\ref{tab:DST_odd_transformations}) in $M_3$ case, that constitute the 3rd AM-QFT variant.

\subsection{The idea behind the 5th, 6th, 7th, 8th QFT variants}

In an amplitude modulation we are not interested in the phase relation between the modulating and modulated signals. 
That is why we can think of employing a sine porting function, instead of a cosine, and expecting to attain the same results of the previous case.
According to this, any already created variant generates a new one, which differs from the original one only for the relation used to convert the odd indices mother signal into even indices child signal:
\begin{itemize}

\item the cosine function is first substituted with the sine one, and specifically:

\begin{itemize}

\item
5th variant: in eq.(\ref{equ:qft_39}) to obtain the basic elaboration $M_5$ from $M_1$

\item
6th variant: in eq.(\ref{equ:qft_53B}) to obtain the basic elaboration $M_6$ from $M_2$

\item
7th variant: in eq.(\ref{equ:qft_59}) to obtain the basic elaboration $M_7$ from $M_3$

\item 
8th variant: in eq.(\ref{equ:qft_46}) to obtain the basic elaboration $M_8$ from $M_4$

\end{itemize}

\item 
the relations in the dual domain, the particular cases, and the involved signal types of these new variants are then obtained accordingly, following the same procedure seen in previous subsections (\emph{mutatis mutandis}) (see Tab.\ref{tab:DCT_odd_transformations},\ref{tab:DST_odd_transformations}).

\end{itemize}
The substitution of the cosine with sine does not affect the computational cost, and the memory requirements, of the new variants.
%La sola differenza rilevante è che, impiegando i seni al posto dei coseni, per delle ragioni trigonometriche, all'atto della trasformazione degli indici dispari negli indici pari, il calcolo di una $\DCT$ viene ricondotto al calcolo di una $\DST$, e viceversa.
%Da ciò consegue anche che cambiano le tipologie dei segnali create, e dunque da gestire.

\begin{figure}[htb]
  \centering
  \includegraphics[width=0.8 \textwidth]{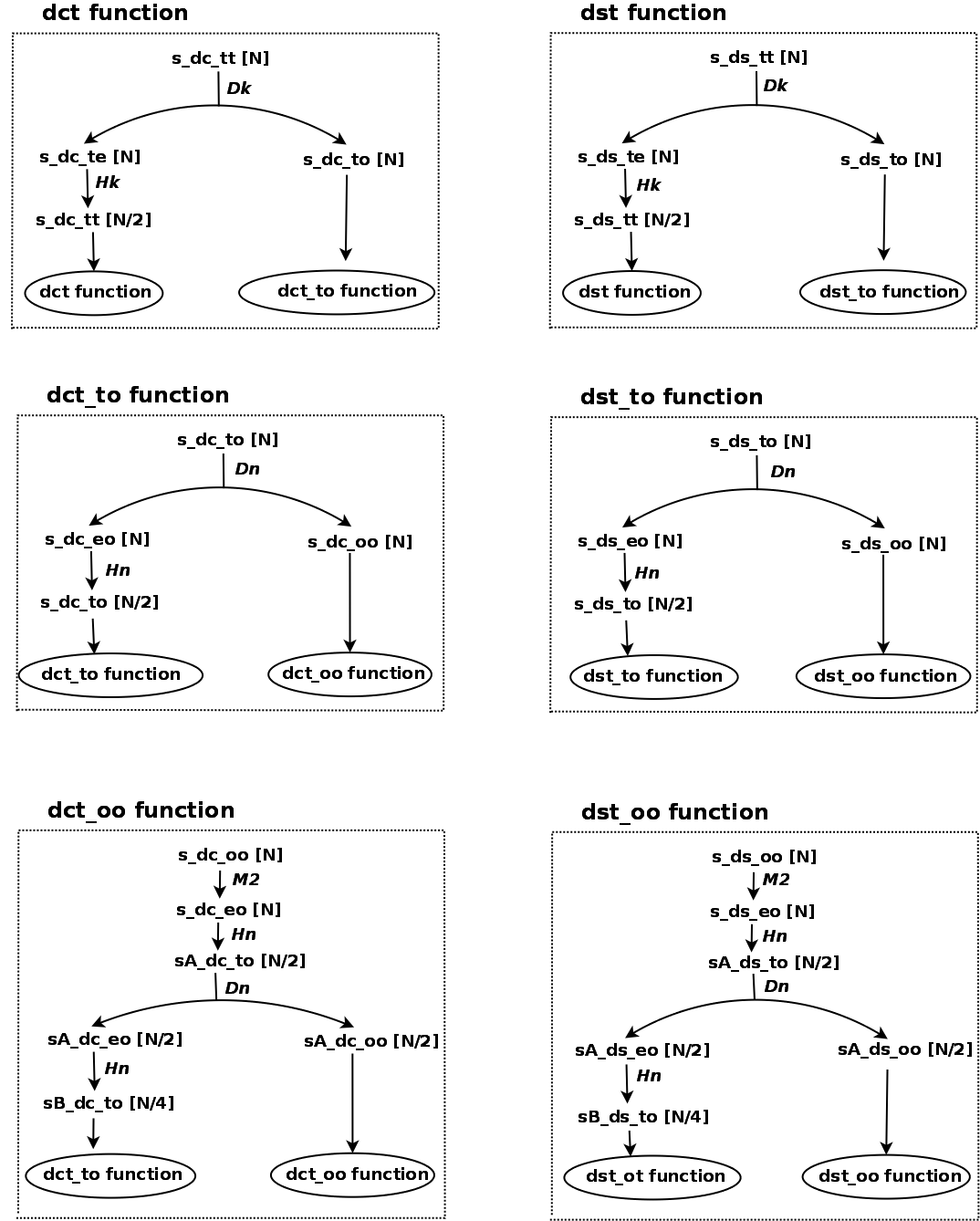}
  \caption{The diagrams of functions used in the 2nd QFT variant}
  \label{fig:QFT_2_diagram}
\end{figure}

% \end{comment}

\begin{figure}[htb]
  \centering
  \includegraphics[width=0.8 \textwidth]{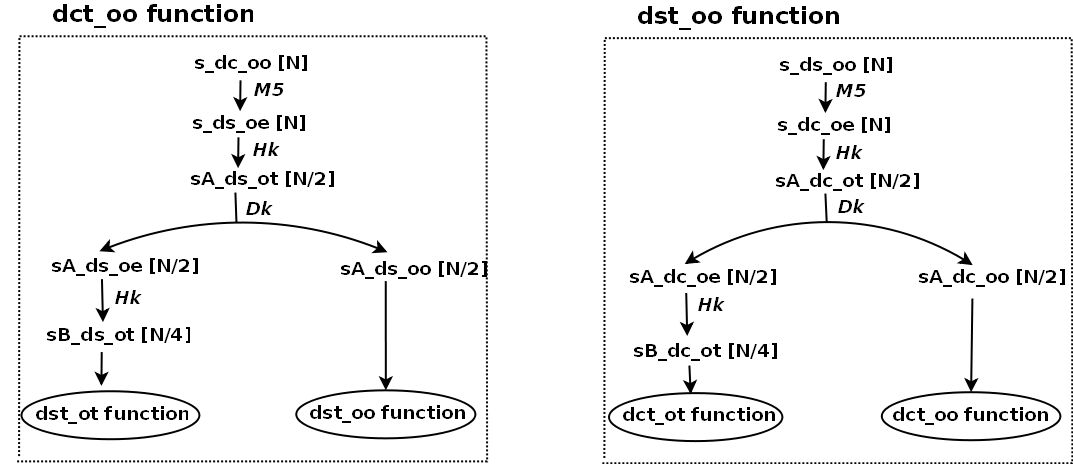}
  \caption{The diagrams of functions used in 5th QFT variant}
  \label{fig:QFT_5_diagram}
\end{figure}

\begin{figure}[htb]
  \centering
  \includegraphics[width=0.8 \textwidth]{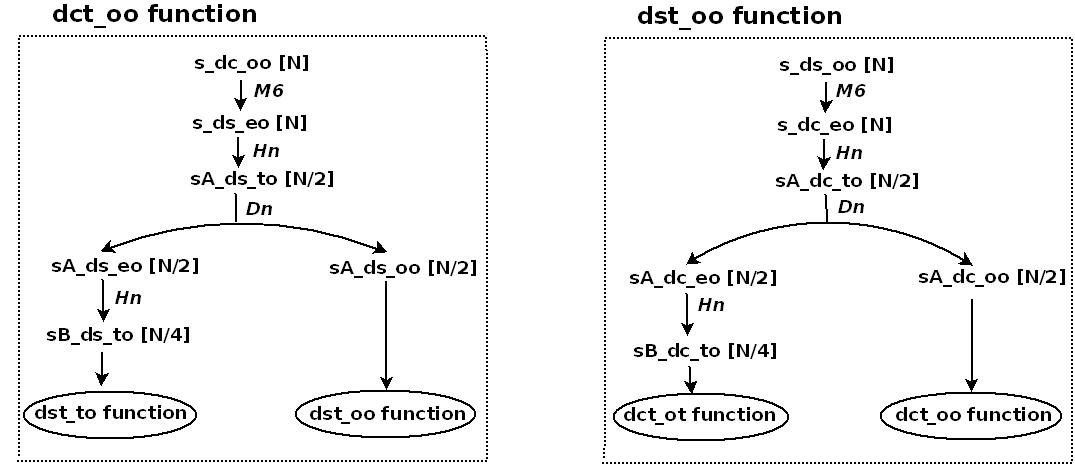}
  \caption{The diagrams of functions used in 6th QFT variant}
  \label{fig:QFT_6_diagram}
\end{figure}

\begin{table}[tb]
\caption{$\DCT[s_{dc\_oo}]$ context: relations involved in transformation of odd indices signal, into even indices signal, in the 8 AM-QFT variants}
\label{tab:DCT_odd_transformations}
\centering
\scalebox{0.60}
{
\begin{tabular}{ccc}
\toprule
& \multicolumn{2}{c}{relations between signals}  \\
\cmidrule{2-3}
basic & temporal relation in DCT context & DCT-frequency domain relation  \\
elaboration & & \\
\midrule

M1 &  $s_{ds\_oe}(n) =  s_{ds \_oo}(n) \cdot 2 \cdot \cos(\theta \cdot n)$  & $\DCT[s_{dc\_oo}](k+1) = \DCT[s_{dc\_oe}](k)- \DCT[s_{dc\_oo}](k-1) $  \\
& $\quad n \in  \{ 1,3,5,\dots, (\frac{N}{4}-1) \}$ & $\quad k \in \{ 2,4,6,\dots, (\frac{N}{4}-2) \}$ \\
&  & $\DCT[s_{dc\_oo}](k=1)=\frac{1}{2} \cdot \DCT[s_{dc \_oe}](k=0)$ \\
 \\

M2 &  $s_{dc\_eo}(n=0) = s_{dc\_oo}(n=1)$ & \\
& $s_{dc\_eo}(n) = s_{dc\_oo}(n+1) + s_{dc\_oo}(n-1)$ & $\DCT[s_{dc\_oo}](k) = \DCT[s_{dc\_eo}](k) \cdot \frac{1}{2 \cdot \cos(\theta \cdot k)}$ \\
& $n \in \{2,4,6,\dots, (\frac{N}{4}-2) \}$ & $k \in \{ 1,3,5,\dots,(\frac{N}{4}-1)\}$ \\
\\

M3 & $s_{dc\_eo}(n=\frac{N}{4}-2) = s_{dc\_oo}(\frac{N}{4}-1)$ & \\
& $s_{dc\_eo}(n) = s_{dc\_oo}(n+1) - s_{dc\_eo}(n+2)$ &  $\DCT[s_{dc\_oo}](k) = \DCT[s_{dc\_eo}](k) \cdot 2 \cdot \cos(\theta \cdot k)$  \\
& $n \in \{2,4,6,\dots, (\frac{N}{4}-4) \}$ & $k \in \{ 1,3,5,\dots,(\frac{N}{4}-1) \}$  \\
& $s_{dc\_eo}(n=0) = \frac{1}{2} \cdot [s_{dc\_oo}(n=1)-s_{dc\_eo}(n=2)]$ &  \\
\\

M4 & $s_{dc \_oe}(n) =  s_{dc\_oo}(n) \cdot \frac{1}{2 \cdot \cos(\theta \cdot n)}$  &  $\DCT[s_{dc\_oo}](k) = \DCT[s_{dc \_oe}](k-1)+ \DCT[s_{dc\_oe}](k+1) $ \\
& $n \in \{ 1,3,5,\dots, (\frac{N}{4}-1) \}$  & $k \in \{ 1,3,5, \dots, (\frac{N}{4}-3) \}$  \\
&  &  $\DCT[s_{dc\_oo}](k=\frac{N}{4}-1)=\DCT[s_{dc\_oe}](k=\frac{N}{4}-2)$ \\

\\

M5 &  $s_{ds\_oe}(n) =  s_{dc\_oo}(n) \cdot 2 \cdot \sin(\theta \cdot n)$  & $\DCT[s_{dc\_oo}](k-1) = \DST[s_{ds\_oe}](k) + \DCT[s_{dc\_oo}](k+1) $  \\
& $\quad n \in  \{ 1,3,5,\dots, (\frac{N}{4}-1) \}$ & $\quad k \in \{ 2,4,6,\dots, (\frac{N}{4}-2) \}$ \\
&  & $\DCT[s_{dc\_oo}](k=\frac{N}{4}-1)=\frac{1}{2} \cdot \DST[s_{ds\_oe}](k=\frac{N}{4})$ \\
 \\

M6 &  $s_{ds\_eo}(n=\frac{N}{4}) = s_{dc\_oo}(n=\frac{N}{4}-1)$ & \\
& $s_{ds\_eo}(n) = s_{dc\_oo}(n-1) - s_{dc\_oo}(n+1)$ & $\DCT[s_{dc\_oo}](k) = \DST[s_{ds\_eo}](k) \cdot \frac{1}{2 \cdot \sin(\theta \cdot k)}$ \\
& $n \in \{2,4,6,\dots, (\frac{N}{4}-2) \}$ & $k \in \{ 1,3,5,\dots,(\frac{N}{4}-1)\}$ \\
\\

M7 & $s_{dc\_eo}(n=2) = s_{dc\_oo}(1)$  & \\
& $s_{ds\_eo}(n) = s_{dc\_oo}(n+1) - s_{dc\_eo}(n+2)$ &  $\DCT[s_{dc\_oo}](k) = \DST[s_{ds\_eo}](k) \cdot 2 \cdot \sin(\theta \cdot k)$  \\
& $n \in \{4,6,8,\dots, (\frac{N}{4}-2) \}$ & $k \in \{ 1,3,5,\dots,(\frac{N}{4}-1) \}$  \\
& $s_{dc\_eo}(n=\frac{N}{4}) = \frac{1}{2} \cdot [s_{dc\_oo}(n=\frac{N}{4}-1)+s_{ds\_eo}(n=\frac{N}{4}-2)]$ &  \\
\\

M8 & $s_{ds\_oe}(n) =  s_{dc\_oo}(n) \cdot \frac{1}{2 \cdot \sin(\theta \cdot n)}$  &  $\DCT[s_{dc\_oo}](k) = \DST[s_{ds\_oe}](k+1)- \DST[s_{ds\_oe}](k-1) $ \\
& $n \in \{ 1,3,5,\dots, (\frac{N}{4}-1) \}$  & $k \in \{ 1,3,5, \dots, (\frac{N}{4}-3) \}$  \\
&  &  $\DCT[s_{dc\_oo}](k=1)=\DST[s_{ds\_oe}](k=2)$ \\

\\

\bottomrule
\end{tabular}
} 
\end{table}

%\begin{sidewaystable} 
\begin{table}[tb]
\caption{$\DST[s_{ds\_oo}]$ context: relations involved in transformation of odd indices signal into even indices signal in the 8 AM-QFT variants}
\label{tab:DST_odd_transformations}
\centering
\scalebox{0.60}
{
\begin{tabular}{ccc}
\toprule
& \multicolumn{2}{c}{relations between signals}  \\
\cmidrule{2-3}
basic & temporal relation in DST context & DST-frequency domain relation  \\
elaboration & & \\
\midrule

M1 &  $s_{ds\_oe}(n) =  s_{ds \_oo}(n) \cdot 2 \cdot \cos(\theta \cdot n)$  & $\DST[s_{ds\_oo}](k+1) = \DST[s_{ds\_de}](k)- \DST[s_{ds\_oo}](k+1) $  \\
& $\quad n \in  \{ 1,3,5,\dots, (\frac{N}{4}-1) \}$ & $\quad k \in \{ 2,4,6,\dots, (\frac{N}{4}-2) \}$ \\
&  & $\DST[s_{ds\_oo}](k=\frac{N}{4}-1)= \frac{1}{2} \cdot \DST[s_{ds\_oe}](k=\frac{N}{4})$ \\
 \\

M2 &  $s_{ds\_eo}(n=\frac{N}{4}) = s_{ds\_oo}(n=\frac{N}{4}-1)$ & \\
& $s_{ds\_eo}(n) = s_{ds\_oo}(n+1) + s_{ds\_oo}(n-1)$ & $\DST[s_{ds\_oo}](k) = \DST[s_{ds\_eo}](k) \cdot \frac{1}{2 \cdot \cos(\theta \cdot k)}$ \\
& $n \in \{2,4,6,\dots, (\frac{N}{4}-2) \}$ & $k \in \{ 1,3,5,\dots,(\frac{N}{4}-1)\}$ \\
\\

M3 & $s_{ds\_eo}(n=2) = s_{ds\_oo}(n=1)$ & \\
& $s_{ds\_eo}(n) = s_{ds\_oo}(n-1) - s_{ds\_eo}(n-2)$ &  $\DST[s_{ds\_oo}](k) = \DST[s_{ds\_eo}](k) \cdot 2 \cdot \cos(\theta \cdot k)$  \\
& $n \in \{2,4,6,\dots, (\frac{N}{4}-4) \}$ & $k \in \{ 1,3,5,\dots,(\frac{N}{4}-1) \}$  \\
& $s_{ds\_eo}(n=0) = \frac{1}{2} \cdot [s_{ds\_oo}(n=\frac{N}{4}-1)-s_{ds\_eo}(n=\frac{N}{4}-2)]$ &  \\
\\

M4 & $s_{ds \_oe}(n) =  s_{ds\_oo}(n) \cdot \frac{1}{2 \cdot \cos(\theta \cdot n)}$  &  $\DST[s_{ds\_oo}](k) = \DST[s_{ds\_oe}](k-1) + \DST[s_{ds\_oe}](k+1) $ \\
& $n \in \{ 1,3,5,\dots, (\frac{N}{4}-1) \}$  & $k \in \{ 3,5,7, \dots, (\frac{N}{4}-1) \}$  \\
&  &  $\DST[s_{ds\_oo}](k=1)=\DST[s_{ds\_oe}](k=2)$ \\
\\

M5 &  $s_{dc\_oe}(n) =  s_{ds \_oo}(n) \cdot 2 \cdot \sin(\theta \cdot n)$  & $\DST[s_{ds\_oo}](k+1) = \DCT[s_{dc\_oe}](k)+ \DST[s_{ds\_oo}](k-1) $  \\
& $\quad n \in  \{ 1,3,5,\dots, (\frac{N}{4}-1) \}$ & $\quad k \in \{ 2,4,6,\dots, (\frac{N}{4}-2) \}$ \\
&  & $\DST[s_{ds\_oo}](k=1)= \frac{1}{2} \cdot \DCT[s_{dc\_oe}](k=0)$ \\
\\

M6 &  $s_{dc\_eo}(n=0) = s_{ds\_oo}(n=1)$ & \\
& $s_{dc\_eo}(n) = s_{ds\_oo}(n+1) - s_{ds\_oo}(n-1)$ & $\DST[s_{ds\_oo}](k) = \DCT[s_{dc\_eo}](k) \cdot \frac{1}{2 \cdot \sin(\theta \cdot k)}$ \\
& $n \in \{2,4,6,\dots, (\frac{N}{4}-2) \}$ & $k \in \{ 1,3,5,\dots,(\frac{N}{4}-1)\}$ \\
\\

M7 & $s_{dc\_eo}(n=\frac{N}{4}-2) = s_{ds\_oo}(n=\frac{N}{4}-1)$ & \\
& $s_{dc\_eo}(n) = s_{ds\_oo}(n+1) + s_{dc\_eo}(n+2)$ &  $\DST[s_{ds\_oo}](k) = \DCT[s_{dc\_eo}](k) \cdot 2 \cdot \sin(\theta \cdot k)$  \\
& $n \in \{2,4,6,\dots, (\frac{N}{4}-4) \}$ & $k \in \{ 1,3,5,\dots,(\frac{N}{4}-1) \}$  \\
& $s_{dc\_eo}(n=0) = \frac{1}{2} \cdot [s_{ds\_oo}(n=1)+s_{dc\_eo}(n=2)]$ &  \\
\\

M8 & $s_{dc \_oe}(n) =  s_{ds\_oo}(n) \cdot \frac{1}{2 \cdot \sin(\theta \cdot n)}$  &  $\DST[s_{ds\_oo}](k) = \DCT[s_{dc\_oe}](k-1)- \DCT[s_{dc\_oe}](k+1) $ \\
& $n \in \{ 1,3,5,\dots, (\frac{N}{4}-1) \}$  & $k \in \{ 1,3,5, \dots, (\frac{N}{4}-3) \}$  \\
&  &  $\DST[s_{ds\_oo}](k=\frac{N}{4}-1)=\DCT[s_{dc\_oe}](k=\frac{N}{4}-2)$ \\
\\

\bottomrule
\end{tabular}
} 
\end{table}
%\end{sidewaystable} 

%%%%%%%%%%%%%%%%%%%%%%%%%%%%%%%%

\section{Recursive description of 8 AM-QFT variant algorithms} \label{sec:6_algorithms}

All QFT variants employ the same number of distinct recursive functions to calculate the $\CDFT$ (8 functions) or the $\RDFT$ (7 functions) transforms.

%%%%%%%%%%%%%%%%%%%%%%

\subsection{The 1st AM-QFT variant algorithm}

The $cdft$, $rdft$, $dct$, $dst$, $dct\_ot$, $dst\_ot$ functions of the 1st variant are identical to the homonymous ones of improved QFT, since both variants use the same elaboration diagrams, shown in Fig.\ref{fig:QFT_4_diagram}.
Differently, the functions  $dct\_oo$ and $dst\_oo$ act in a similar way (but are not identical) to the homonymous functions of improved QFT, since 
they use the $M_1$ basic elaboration (described in Tab.\ref{tab:DCT_odd_transformations},\ref{tab:DST_odd_transformations}), instead of the $M_4$ one.

%%%%%%%%%%%%%%%%%%%%%%%%%%%%%%

\subsection{The 2nd AM-QFT variant algorithm}

\begin{figure}[htb]
  \centering
  \includegraphics[width=1.5 \textwidth, angle=90]{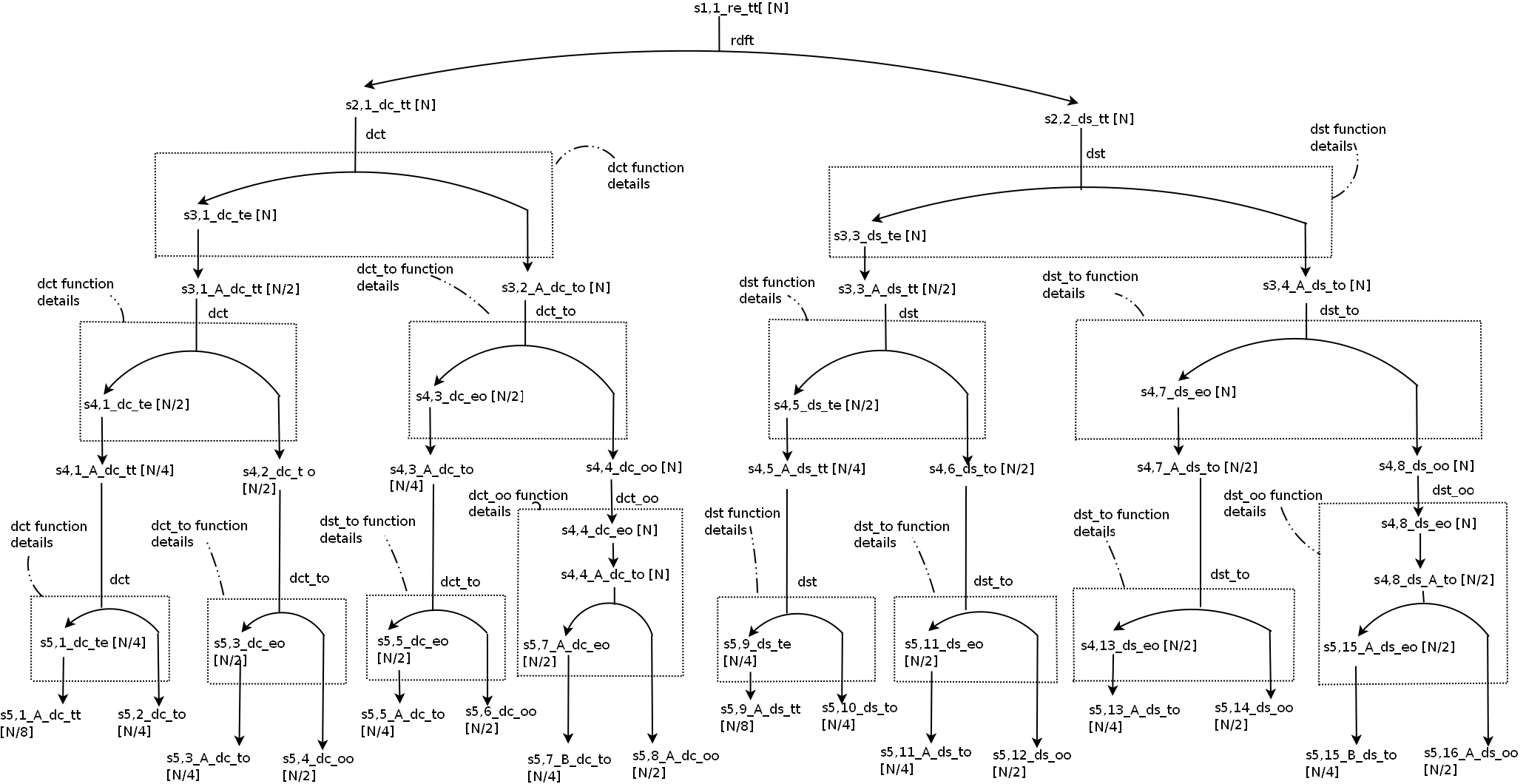}
  \caption{The decomposition tree of the 2nd QFT variant}
  \label{fig:QFT_2_tree}
\end{figure}

The $cdft$, $rdft$, $dct$, $dst$ functions coincide with those employed in improved QFT. The remaining functions $dct\_to$, $dct\_oo$, $dst\_oo$, $dst\_to$ can be developed starting from the diagrams shown in Fig.\ref{fig:QFT_2_diagram} and using Tab.\ref{tab:elabora_relations_A},\ref{tab:elabora_relations_B},\ref{tab:DCT_odd_transformations},\ref{tab:DST_odd_transformations} to convert abstract basic elaborations into temporal and frequency-domain mathematical relations, as shown in sect.\ref{sec:improved_QFT}.
Let us observe that the roles of time and frequency are swapped (both in signal notation and in basic elaborations) with respect to the 1st variant and the improved QFT.
Moreover the concatenation of elaborations diagrams associated to the functions used in this 2nd QFT variant generates the decomposition tree shown in Fig.\ref{fig:QFT_2_tree}.

%%%%%%%%%%%%%%%%%%%%%%%%%%%%%%%%%%%%%%%%

\subsection{The 3rd AM-QFT variant algorithm}

The $cdft$, $rdft$, $dct$, $dst$,  $dct\_to$, $dst\_to$ functions employed in this 3rd variant coincide with those employed in the 2nd variant. 
The remaining functions $dct\_oo$ and $dst\_oo$ can be developed starting from the diagrams shown in Fig.\ref{fig:QFT_2_diagram}, changing the $M_2$ basic elaboration with $M_3$ one, and using Tab.\ref{tab:elabora_relations_A},\ref{tab:elabora_relations_B},\ref{tab:DCT_odd_transformations},\ref{tab:DST_odd_transformations} as shown in sect. \ref{sec:improved_QFT}.

%%%%%%%%%%%%%%%%%%%%%%%%%%%%%%%%%

\subsection{The 4th AM-QFT variant algorithm}

This variant coincide with the improved QFT algorithm \cite{Pasquini_2013} already described in sect. \ref{sec:improved_QFT}.

%%%%%%%%%%%%%%%%

\subsection{The 5th AM-QFT variant algorithm}

The $cdft$, $rdft$, $dct$, $dst$, $dct\_ot$, $dct\_ot$ functions employed in this 5th variant coincide with those employed in improved QFT. The remaining functions $dct\_oo$ and $dst\_oo$ can be developed using the elaboration diagrams shown in Fig.\ref{fig:QFT_5_diagram} and using Tab.\ref{tab:elabora_relations_A}, \ref{tab:elabora_relations_B},\ref{tab:DCT_odd_transformations},\ref{tab:DST_odd_transformations}, as shown in sect.\ref{sec:improved_QFT}.

%%%%%%%%%%%%%%%%

\subsection{The 6th AM-QFT variant algorithm}

The $cdft$, $rdft$, $dct$, $dst$, $dct\_to$, $dct\_to$ functions employed in this 6th variant coincide with those employed in 2nd variant. 
The remaining functions $dct\_to$, $dct\_oo$, $dst\_oo$, $dst\_to$ can be developed using the diagrams shown in Fig.\ref{fig:QFT_6_diagram} and Tab.\ref{tab:elabora_relations_A},\ref{tab:elabora_relations_B},\ref{tab:DCT_odd_transformations},\ref{tab:DST_odd_transformations}, as shown in sect.\ref{sec:improved_QFT}.
Let us observe that in this case, analougously to the 1st/2th variants case, we have again a time/frequency swap with respect to the 5th variant.

%%%%%%%%%%%%%%%%

\subsection{The 7th AM-QFT variant algorithm}

The $cdft$, $rdft$, $dct$, $dst$,  $dct\_to$, $dst\_to$ functions employed in this 7th variant coincide with those employed in 3rd variant. 
The remaining functions $dct\_oo$ and $dst\_oo$ can be developed starting from the diagrams shown in Fig.\ref{fig:QFT_6_diagram}, changing the $M_6$ basic elaboration with $M_7$ and using Tab.\ref{tab:elabora_relations_A},\ref{tab:elabora_relations_B},\ref{tab:DCT_odd_transformations},\ref{tab:DST_odd_transformations}, as shown in sect. \ref{sec:improved_QFT}.

%%%%%%%%%%%%%%%%%%%%%%%%%%%%%%%%%%%%%%%%%%%%%%%%%%%

\subsection{The 8th AM-QFT variant algorithm}

The $cdft$, $rdft$, $dct$, $dst$,  $dct\_ot$, $dst\_ot$ functions employed in this 8th variant coincide with those employed in 4th variant. 
The remaining functions $dct\_oo$ and $dst\_oo$ can be developed starting from the diagrams shown in Fig.\ref{fig:QFT_5_diagram}, changing the $M_5$ basic elaboration with $M_8$, and using Tab.\ref{tab:elabora_relations_A},\ref{tab:elabora_relations_B},\ref{tab:DCT_odd_transformations},\ref{tab:DST_odd_transformations}, as shown in sect. \ref{sec:improved_QFT}.

\subsection{General notes on AM-QFT variants}

The main difference between the first four variants versus the other ones, is that the last ones mixes $\DCT$ and $\DST$ contexts, since the computations of $\DCT$ is transformed into the computation of a $\DST$ and viceversa.
It follows that the computation of $\DCT-0$ or $\DST-0$ transforms requires three functions using the first four variants, and five functions using the remaining variants.
Moreover it must be observed that, in each variant, the even/odd separation of time indices can be performed both before and after the even/odd separation of harmonics. 
In this regard we have choosen the order that minimizes the number of distinct involved functions. 
Thus in the 1st, 4th, 5th and 8th algorithm variants we first separate the temporal indices and then the frequency-domain ones, and viceversa in the remaining variants.
At the light of these rules, in any variant the transformation of odd indices into the even ones is applied only to signal types $s_{dc\_oo}$ and $s_{ds \_oo}$.

%%%%%%%%%%%%%%%%%%%%%%%%%%%%%%%%

\section{The characteristics of 8 variants of QFT} \label{sec:costo_computazionale}

%\label{sec:memoria_richiesta}

\subsection{Memory Requirements}
The eight AM-QFT variants require the same amount of $\frac{N}{4}$ distinct real trigonometric constants (used only in $dct\_oo$ and $dst\_oo$ functions).
The constant $cos(\frac{2 \dot \pi}{8})=sin(\frac{2 \dot \pi}{8})$, that is used in the special case $N=8$ of $dct\_oo$ and $dst\_oo$ functions, is common to any variant.
In the not `on the fly' implementation case, the remaining $\frac{N}{4}-1$ trigonometric constants that we need to store and to a-priori calculate, are of type: 
$2 \cdot \cos(\theta \cdot p) \quad p \in \{ 1,2,3, \dots,(\frac{N}{4}-1) \}$ in the 1st and 3rd variant,
$2 \cdot \sin(\theta \cdot p) \quad p \in \{ 1,2,3, \dots,(\frac{N}{4}-1) \}$ in the 5th and 7th variant,
$\frac{1}{2 \cdot \cos(\theta \cdot p)} \quad p \in \{ 1,2,3, \dots,(\frac{N}{4}-1) \}$ in the 2nd and 4th variants,
$\frac{1}{2 \cdot \sin(\theta \cdot p)} \quad p \in \{ 1,2,3, \dots,(\frac{N}{4}-1) \}$ in the 6th and 8th variants.
It is easy to observe that the subclass of 1st, 3rd, 5th and 7th variants employ the same trigonometric constants set, and the same holds for the subclass of 2nd, 4th, 6th, 8th variants, since the sequence of sines is equivalent to the sequence of cosines in reverse order, and the same applies for secant/cosecant relationship.
All variants (as well as for the split-radix and the tangent FFT \cite{Bernstein_2007}) can be implemented in-place too (differently from classical QFT \cite{Guo_Sitton_qft_1998} that can be in-place only if the goal is the $\DST$ computation, not for $\DCT$ or $\DFT$ computation).
The reason is that any employed function in AM-QFT class leaves unchanged the total number of temporal and frequency-domain elements to be stored, uses a fixed number of inner temporary variables (not depending on periodization $N$), and uses only intrinsecally implementable in-place basic elaboration (if handled in an isolated way, not depending in input/output indices order).
However an efficient (with a few data moves) in-place implementation of this AM-QFT class requires future work.

%%%%%%%%%%%%%%%%%%%%%%%%%%%%%%%%%%%%%%%%%%%%%%%%%%%

%\section{Computational cost of the proposed QFT variants} \label{sec:costo_computazionale}
\subsection{Computational Cost}

Tab.\ref{tab:costo_varianti},\ref{tab:flop} describe the computational cost of the class of AM-QFT algorithms (as usual, this evaluation is referred to not `on the fly' algorithm implementation, that is the calculation  of trigonometric constants $2 \cdot \cos(\theta \cdot n)$, $2 \cdot \sin(\theta \cdot n)$, $\frac{1}{2 \cdot \cos(\theta \cdot n)}$, $\frac{1}{2 \cdot \sin(\theta \cdot n)}$ have been performed a-priori). 
We have already pointed out that the 1st, 3rd, 5th and 7th variants require also some divisions by two, and specifically in operations related to transformations of odd indices in even ones (shown in Tab.\ref{tab:DCT_odd_transformations} and in Tab.\ref{tab:DST_odd_transformations}) for $M_1$, $M_3$, $M_5$, $M_7$ cases. 
The computational burden associated to such operation, both in HW and SW case, is typically less than a generic multiplication, specially in fixed-point implementation (assuming to use a binary representation for numbers). 
Thus we decide not to include the binary translations into the multiplications account, but to consider them separately. 
Moreover we evaluate the algorithm flop requirements both with and without considering such binary translations.
If we neglect the binary translations, then any AM-QFT variant requires the same sums, multiplications, flops counts.
Moreover these counts are identical to split-radix 3mul-3add and improved QFT cases.
Differently, if we insert the binary translations into the flop count, then only the 2nd, 4th, 6th, 8th variants require the same flop counts. 
Moreover, among the algorithms addressed in Tab.\ref{tab:compara_add},\ref{tab:compara_mol},\ref{tab:compara_flop}, the split-radix 3add/3mul and the QFT variants require the least number of multiplications.
These theoretical results are confirmed by a toy algorithm implemented in Scilab environment, that counts all the arithmetical operations for each called function.

%%%%%%%%%%%%%%%%

\subsection{Accuracy}

\begin{figure}[htb]
  \centering
  \includegraphics[width=0.6 \textwidth]{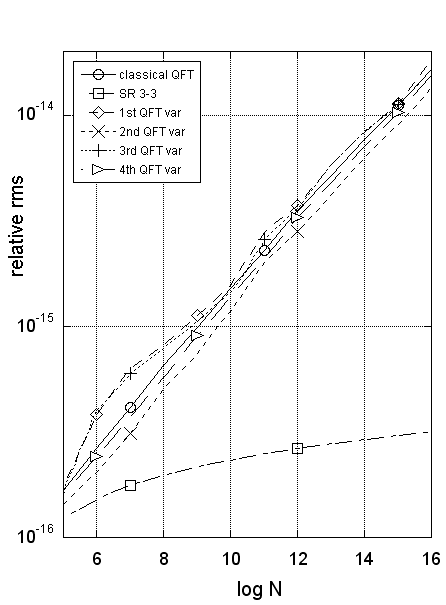}
  \caption{The accuracy of the 1st 2nd, 3rd and 4th variants of AM-QFT. Legend: SR 3-3= Split-Radix 3mul-3add}
  \label{fig:accuracy_1}
\end{figure}

\begin{figure}[htb]
  \centering
  \includegraphics[width=0.6 \textwidth]{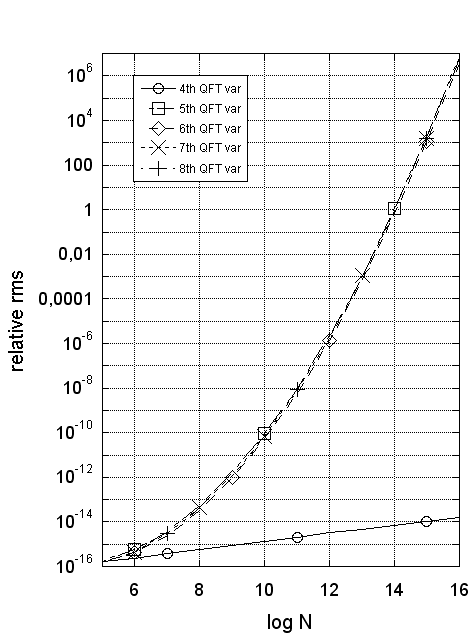}
  \caption{The accuracy of 5th 6th, 7th and 8th variants of AM-QFT}
  \label{fig:accuracy_2}
\end{figure}

The accuracy of 8 variants of AM-QFT algorithm is reported in Fig.\ref{fig:accuracy_1},\ref{fig:accuracy_2}.
We surprisingly note that the numerical error of the 5th, 6th, 7th and 8th variants (that use sine function) grows far faster with respect to the one of the other variants  (that use cosine function).
Curiously, comparing Fig.\ref{fig:accuracy_2} with graphs in \cite{Johnson_Frigo_2000}, we can argue that the 5th, 6th, 7th, 8th variants of AM-QFT class are the worst accurate FFT algorithms ever published!
Fig.\ref{fig:accuracy_1} shows that the 2nd variant is the most accurate in AM-QFT class.
In many applications the not excellent accuracy of 1st, nd, 3rd, 4th QFT variants (if compared to split-radix) is not very important, since we are interested only to few digits of frequency-domain signals values, and thus obtaining a relative error about $10^{-14}$ or $10^{-16}$ is quite the same. 
Let us observe that the 1st and 3rd variants are less accurate than the 2nd variant, also if they use the cosine trigonometric constants array (that is much more accurate than the secant array, both as absolute error, and as relative error).
We explain the reason only for the 1st variant in $\DCT$ context (the $\DST$ context and the 3rd variant cases are analogous).
The $M_1$ basic elaboration, in $\DCT$ context (see Tab.\ref{tab:DCT_odd_transformations}) forces us to compute $S_{dc\_oo}(k+1)$ value using the previously computed $S_{dc\_oo}(k-1)$ value of the same signal, for any $k \in sto\_k$. 
As a result, the last computed value $S_{dc\_oo}(k=\frac{N}{4}-1)$ is far less accurate with respect to the first computed value $S_{dc\_oo}(k=1)$ of the same signal, because of cumulation of errors due to this recursive process required by $M_1$ basic elaboration.
Differently this phenomena of cumulation of error does not happen in the 2nd or 4th variant, where we use $M_2$, $M_4$ respectively, instead of $M_1$, $M_3$ basic elaborations.

\begin{table}[tb]
\caption{Computational cost required for various sinusoidal transforms by means of the proposed QFT variants, in dependence on their periodization $N$. The binary translations are only required for the 1st, 3rd, 5th and 7th QFT variants}
\label{tab:costo_varianti}
\centering
\scalebox{0.8}
{
\begin{tabular}{cccc}
\toprule
& \multicolumn{3}{c}{computational cost}  \\
\cmidrule{2-4}
transform & multiplications & sums & binary translations \\
\midrule
CDFT & $N \log(N)-3 N+4$ & $3 N \log(N)-3 N+4$ & $N-4 \log(N)+4$  \\
RDFT & $\frac{1}{2} N \log(N)- \frac{3}{2} N+2$ & $\frac{3}{2} N \log(N)-\frac{5}{2} N+4$ & $\frac{1}{2} N-2 \log(N)+2$  \\
DCT & $\frac{1}{4} N \log(N)- \frac{3}{4} N+1$ & $\frac{3}{4} N \log(N)-\frac{7}{4} N+\log(N)+3$ & $\frac{1}{4} N-\log(N)+1$  \\
DST & $\frac{1}{4} N \log(N)- \frac{3}{4} N+1$ & $\frac{3}{4} N \log(N)-\frac{7}{4} N-\log(N)+3$ & $\frac{1}{4} N-\log(N)+1$  \\
\bottomrule
\end{tabular}
} 
\end{table}

\begin{table}[tb]
\small
\caption{Number of flops required to calculate different sinusoidal transforms by means of the proposed QFT variants, in dependence on their periodization $N$. 
The case A refers to the 2nd, 4th, 6th,8th variants subclass (and to the 1st, 3rd, 5th,7th variants  subclass too, if we neglect the binary translations).
The case B refers to the 1st, 3rd, 5th,7th variants subclass, if we insert the binary translations into the flop count}
\label{tab:flop}
\centering
\scalebox{0.8}
{
\begin{tabular}{ccc}
\toprule
trasnform & flop (case A) & flop (case B) \\
\midrule
CDFT & $4 N \log(N)-6 N+8$ & $4 N \log(N)-5 N-4 \log(N)+12$ \\
RDFT & $2 N \log(N)-4 N+6$ & $2 N \log(N)-\frac{7}{2} N-2 \log(N)+8$ \\
DCT & $N \log(N)- \frac{5}{2} N+\log(N)+4$ & $N \log(N)-\frac{9}{4} N +5$ \\
DST & $N \log(N)- \frac{5}{2} N-\log(N)+4$ & $N \log(N)-\frac{9}{4} N - 2 \log(N) +5$ \\
\bottomrule
\end{tabular}
}
\end{table}

\begin{table}[tb]
\caption{Comparative evaluation of number of sums required for $\CDFT$ calculation with various algorithms. 
Legend: var\_QFT = QFT\_variants, SR\_4-2=Split-Radix 4mul-2add, SR\_3-3=Split-Radix 3add-3mul,  JF= scaled split-radix by Johnson and Frigo, clas\_QFT = classical\_QFT }
\label{tab:compara_add}
\centering
\scalebox{0.8}
{
\begin{tabular}{cccccc}
\toprule
& \multicolumn{5}{c}{sums}  \\
\cmidrule{2-6}
N & $var\_QFT$ & $SR\_4/2$ & $SR\_3/3$ & $JF$ & $clas\_QFT$ \\
\midrule
$4$ & $16$ & $16$ & $16$ & $16$ & $16$ \\
$8$ & $52$ & $52$ & $52$ & $52$ & $52$ \\
$16$ & $148$ & $144$ & $148$ & $144$ & $160$ \\
$32$ & $388$ & $372$ & $388$ & $372$ & $432$ \\
$64$ & $964$ & $912$ & $964$ & $912$ & $1088$ \\
$128$ & $2308$ & $2164$ & $2308$ & $2164$ & $2624$ \\
$256$ & $5380$ & $5008$ & $5380$ & $5008$ & $6144$ \\
$512$ & $12292$ & $11380$ & $12290$ & $11380$ & $14080$ \\
$1024$ & $27652$ & $25488$ & $27652$ & $25488$ & $31744$ \\
$2048$ & $61444$ & $56436$ & $61444$ & $56436$ & $70656$ \\
\bottomrule
\end{tabular}
}
\end{table}

\begin{table}[tb]
\caption{Comparative evaluation of number of multiplications required for $\CDFT$ calculation with various algorithms. 
Legend: var\_QFT = QFT\_variants, SR\_4-2=Split Radix 4mul-2add, SR 3-3=Split Radix 3add-3mul,  JF=scaled split radix by  Johnson and Frigo, clas\_QFT = classical\_QFT }
\label{tab:compara_mol}
\centering
\scalebox{0.8}
{
\begin{tabular}{cccccc}
\toprule
& \multicolumn{5}{c}{multiplications}  \\
\cmidrule{2-6}
N & $var\_QFT$ & $SR\_4/2$ & $SR\_3/3$ & $JF$ & $QFT\_clas$ \\
\midrule
$4$ & $0$ & $0$ & $0$ & $0$ & $0$ \\
$8$ & $4$ & $4$ & $4$ & $4$ & $4$ \\
$16$ & $20$ & $24$ & $20$ & $24$ & $22$ \\
$32$ & $68$ & $84$ & $68$ & $84$ & $74$ \\
$64$ & $196$ & $248$ & $196$ & $240$ & $210$ \\
$128$ & $516$ & $660$ & $516$ & $628$ & $546$ \\
$256$ & $1284$ & $1656$ & $1284$ & $1544$ & $1346$ \\
$512$ & $3076$ & $3988$ & $3076$ & $3668$ & $3202$ \\
$1024$ & $7172$ & $9336$ & $7172$ & $8480$ & $7426$ \\
$2048$ & $16388$ & $21396$ & $16388$ & $19252$ & $16898$ \\
\bottomrule
\end{tabular}
}
\end{table}

\begin{table}[tb]
\caption{Comparative evaluation of number of flops required for $\CDFT$ calculation with various algorithms. 
Legend: var\_QFT = QFT\_variants, SR = Split Radix, JF = scaled split radix by Johnson and Frigo, clas\_QFT = classical\_QFT. 
The case A refers to the 2nd, 4th, 6th,8th variants subclass (and to the 1st, 3rd, 5th,7th variants  subclass too, if we neglect the binary translations).
The case B refers to the 1st, 3rd, 5th,7th variants subclass, if we insert the binary translations into the flop count }
\label{tab:compara_flop}
\centering
\scalebox{0.8}
{
\begin{tabular}{cccccc}
\toprule
& \multicolumn{5}{c}{flop}  \\
\cmidrule{2-6}
N & $var\_QFT$ & $var\_QFT$ & $SR$ & $JF$ & $clas\_QFT$ \\
 & case A  & case B & & & \\
\midrule
$4$ & $16$ & $16$ & $16$ & $16$ & $16$ \\
$8$ & $56$ & $56$ & $56$ & $56$ & $56$ \\
$16$ & $168$ & $172$ & $168$ & $168$ & $182$ \\
$32$ & $456$ & $472$ & $456$ & $456$ & $506$ \\
$64$ & $1160$ & $1204$ & $1160$ & $1152$ & $1298$ \\
$128$ & $2824$ & $2928$ & $2824$ & $2792$ & $3170$ \\
$256$ & $6664$ & $6892$ & $6664$ & $6552$ & $7490$ \\
$512$ & $15368$ & $15848$ & $15368$ & $15048$ & $17282$ \\
$1024$ & $34824$ & $35812$ & $34824$ & $33968$ & $39170$ \\
$2048$ & $77832$ & $79840$ & $77832$ & $75688$ & $87554$ \\
\bottomrule
\end{tabular}
}
\end{table}

%%%%%%%%%%%%%%%%%%%%%%%%%%%%%%%%%%%%%%%%%%%%

%%%%%%%%%%%%%%%%%%%%%%%%%%%%%%%%%%%%

\subsection{Applications of QFT variants}

The 1st, 2nd, 3rd and 4th variants cover the whole range of possible applications of FFT algorithms.
In fact the 2nd and 4th variants (the latter being the already published improved QFT) are suitable for `not on the fly' implementations.
On the contrary the the 1st and 3rd variants are the proper choise in the `on the fly' context, by virtue of simplicity of their trigonometric constants.
Thus the most competitive algorithm to which the proposed QFT variants can be compared with is the split-radix. 
To be more precise the main applications are:
\begin{itemize}

\item 
multiple sinusoidal transforms ($\CDFT$, $\RDFT$, $\DCT-0$, $\DST-0$) computation in SW environments like SCILAB, MATLAB or MAPLE, running on PC platforms. Indeed, within these environments, the user typically requires the `on the fly' calculation of a single transform applied to a certain signal. 
In this context, at difference with split-radix, we just need to write, optimize and memorize only a piece of code to calculate all the above different transforms.

\item
Fixed-point implementation both `on the fly' and not `on the fly', due to the low number of multiplications and the few simple trigonometric constants to calculate. 
For example the implementation on low-cost DSP or MPU, with scarce computational resources (wihout floating-point arithmetic), is particularly recommended.

\item
parallel pipeline hardware implementation.

\end{itemize}

%%%%%%%%%%%%%%%%%%%%%%%%%%%%%%%%%%%%%%%%%%%%%%

\section{Conclusions}
\label{sec:finale}

We can summarize the work outcomes saying that we have obtained a class of 8 AM-QFT variants that are more accurate, or with faster trigonometric constants in on the fly implementation, then improved QFT.
Moreover, in certain applicative contexts, some variants have more attractive properties with respect to the split-radix 3mul-3add algorithm, since they require the same multiplications, additions and flops, but with half of the trigonometric constants. 
In our opinion the proposed approach represents one of the best compromise in achieving the quality standards typically required to an FFT algorithm.
Finally the approach used in this paper seems to be particularly fit to describe other popular FFT algorithms, such as radix-2, radix-4 and split-radix.

%%%%%%%%%%%%%%%%%%%%%%%%%%%%%%%%%%%%%%%

\section{Acknoledgments}

Michele Pasquini, Stefano Squartini and Francesco Piazza helped the author in revision and translation of this paper.

%%%%%%%%%%%%%%%%%%%%%%%

\bibliographystyle{plain}

\bibliography{biblio}

\begin{thebibliography}{10}

\bibitem{Bernstein_2007}
Daniel~J. Bernstein.
\newblock The tangent fft.
\newblock In {\em Boztas and Lu}, pages 291--300, 2007.

\bibitem{Bouguezel_2007}
Saad Bouguezel, M.~Omair Ahmad, and M.~N.~S. Swamy.
\newblock A general class of split-radix fft algorithms for the computation of
  the dft of length-2$^{\mbox{m}}$.
\newblock {\em IEEE Transactions on Signal Processing}, 55(8):4127--4138, 2007.

\bibitem{Cho_Themes_1978}
K.~M. Cho and G.~C. Themes.
\newblock Real-factor fft algorithms.
\newblock {\em IEEE}, 1978.

\bibitem{Duhamel_Vetterli_1990}
P.~Duhamel and M.~Vetterli.
\newblock Fast fourier transforms: a tutorial review and a state of the art.
\newblock {\em Signal Process.}, 19:259--299, 1990.

\bibitem{Duhamel_Hollmann_1984}
Pierre Duhamel and H.~Hollmann.
\newblock {Split-radix FFT algorithm}.
\newblock {\em Electronics Letters}, 20:14--16, 1984.

\bibitem{Gopinath_1989}
Gopinath.
\newblock Comment conjugate pair fast fourier transform.
\newblock {\em Electronics Letters}, 25(16):1084, 1989.

\bibitem{Guo_Sitton_qft_1998}
Haitao Guo, Gary~A. Sitton, and C.~Sidney Burrus.
\newblock The quick fourier transform: an fft based on symmetries.
\newblock {\em IEEE Transactions on Signal Processing}, 46(2):335--341, 1998.

\bibitem{Johnson_Frigo_2007}
Steven~G. Johnson and Matteo Frigo.
\newblock A modified split-radix fft with fewer arithmetic operations.
\newblock {\em IEEE Transactions on Signal Processing}, 55(1):111--119, 2007.

\bibitem{Kamar_Elcherif_1989}
I.~Kamar and Y.~Elcherif.
\newblock Conjugate pair fast fourier transform.
\newblock {\em Electronics Letters}, 25(5):324--325, 1989.

\bibitem{Lundy_Van_Buskirk_2007}
T.~Lundy and J.~Van Buskirk.
\newblock A new matrix approach to real ffts and convolutions of length $2^{
  k}$.
\newblock {\em Computing}, 80(1):23--45, 2007.

\bibitem{Martens_1984}
Jean-Bernard Martens.
\newblock {Recursive cyclotomic factorization---a new algorithm for calculating
  the discrete Fourier transform}.
\newblock {\em IEEE Transactions on Acoustics, Speech, and Signal Processing},
  32:750--761, 1984.

\bibitem{Pasquini_2013}
Lorenzo Pasquini.
\newblock Improved qft algorithm for power-of-two fft.
\newblock {\em arxiv.org (pre-print)}, 2013.

\bibitem{Johnson_Frigo_2000}
M.~Frigo S.~Johnson.
\newblock (online) http://www.fftw.org/accuracy/.

\bibitem{Stasinski_1991}
Ryszard Stasinski.
\newblock {The techniques of the generalized fast Fourier transform algorithm}.
\newblock {\em IEEE Transactions on Signal Processing}, 39:1058--1069, 1991.

\bibitem{Vetterli_Nussbaumer_1984}
Martin Vetterli and Henri~J. Nussbaumer.
\newblock Simple {FFT} and {DCT} algorithms with reduced number of operations.
\newblock {\em Signal Processing}, 6(4):267--278, 1984.

\bibitem{Yavne_1968}
R.~Yavne.
\newblock {An economical method for calculating the discrete Fourier
  transform}.
\newblock In {\em AFIPS '68 (Fall, part I): Proceedings of the joint computer
  conference}, pages 115--125, New York, NY, USA, 1968. ACM.

\end{thebibliography}

\end{document}